\documentclass[
 aip,
 long, numerical bibliography, (default)
 jmp,%
 amsmath,amssymb,
preprint,%
nofootinbib
]{revtex4-1}

\usepackage[a4paper,margin=2.5cm]{geometry}
\usepackage[utf8]{inputenc}

\usepackage{graphicx}% Include figure files
\usepackage{dcolumn}% Align table columns on decimal point
\usepackage{bm}% bold math
\usepackage{epstopdf}
\usepackage{xr}
\usepackage{amsmath}
\usepackage{color}
\usepackage[normalem]{ulem}
\usepackage{wasysym}
\usepackage{siunitx}
\usepackage{physics}
\usepackage{textcomp}
\usepackage{xcolor}

\makeatletter

\newcommand{\Rmnum}[1]{\expandafter\@slowromancap\romannumeral #1@}
\makeatother

\begin{document}
%\markboth{\jobname}{\jobname .tex}
%\preprint{AIP/123-QED}

\title{Landau level diagram and the continuous rotational symmetry breaking in trilayer graphene}
\author{Biswajit Datta}
\affiliation{Department of Condensed Matter Physics and Materials Science, Tata Institute of Fundamental Research, Homi Bhabha Road, Mumbai 400005, India}
\author{Hitesh Agarwal}
\affiliation{Department of Condensed Matter Physics and Materials Science, Tata Institute of Fundamental Research, Homi Bhabha Road, Mumbai 400005, India}
\author{Abhisek Samanta}
\affiliation{Department of Theoretical Physics, Tata Institute of Fundamental Research, Homi Bhabha Road, Mumbai 400005, India}
\author{Amulya Ratnakar}
\affiliation{Department of Physics, UM-DAE Centre for Excellence in Basic Sciences, Mumbai  400098, India}
\author{Kenji Watanabe}
\affiliation{National Institute for Materials Science, 1-1 Namiki, Tsukuba 305-0044, Japan}
\author{Takashi Taniguchi}
\affiliation{National Institute for Materials Science, 1-1 Namiki, Tsukuba 305-0044, Japan}
\author{Rajdeep Sensarma}
\homepage{sensarma@theory.tifr.res.in}
\affiliation{Department of Theoretical Physics, Tata Institute of Fundamental Research, Homi Bhabha Road, Mumbai 400005, India}
\author{Mandar M. Deshmukh}
\homepage{deshmukh@tifr.res.in}
\affiliation{Department of Condensed Matter Physics and Materials Science, Tata Institute of Fundamental Research, Homi Bhabha Road, Mumbai 400005, India}
%\date{\today}

%abstract

\begin{abstract}

The sequence of the zeroth Landau levels (LLs) between filling factors $\nu$=-6 to 6 in ABA-stacked trilayer graphene (TLG) is unknown because it depends sensitively on the non-uniform charge distribution on the three layers of ABA-stacked TLG. Using the sensitivity of quantum Hall data on the electric field and magnetic field, in an ultraclean ABA-stacked TLG sample, we quantitatively estimate the non-uniformity of the electric field and determine the sequence of the zeroth LLs. We also observe anticrossings between some LLs differing by 3 in LL index, which result from the breaking of the continuous rotational to \textit{C}$_3$ symmetry by the trigonal warping.

\end{abstract}

\maketitle

After extensive studies on graphene~\cite{novoselov_two-dimensional_2005,castro_neto_electronic_2009} and bilayer graphene~\cite{mccann_electronic_2013,Guinea_2006,Latil_2006,Partoens_2006}, TLG has emerged as a new platform to study interactions between Dirac electrons in a controlled manner. The two variants, namely Bernal (ABA)~\cite{datta_strong_2017,inplane_B,Yuan_2011,henriksen_quantum_2012,craciun2009trilayer,campos_landau_2016,taychatanapat_quantum_2011,lee2013broken,zhao_symmetry_2010,stepanov_tunable_2016,mccann_landau-level_2006,zhang2012hund} and rhombohedral (ABC)~\cite{lee_competition_2014,zou_transport_2013,bao_stacking-dependent_2011,kumar_integer_2011,Jhang_2011,lui_observation_2011,Elferen_2013,lui2010imaging} stackings have been studied for their tunable symmetries. ABA-stacked TLG has a rich low energy band-structure consisting of a monolayer graphene (MLG)-like linear and a bilayer graphene (BLG)-like quadratic bands~\cite{serbyn_new_2013,koshino_gate-induced_2009}. Further, the mirror symmetry in ABA-stacked TLG, inhibits any coupling between the linear and quadratic bands~\cite{koshino_landau_2011,serbyn_new_2013}. In the presence of a transverse electric field (\textit{E}$^\perp$), however, the mirror symmetry is broken, resulting in the mixing of wavefunctions from two subbands, and thus providing a tunable knob to probe the interesting physics resulting from the mixing of wavefunctions from the linear and quadratic bands~\cite{koshino_gate-induced_2009,serbyn_new_2013}.

Trigonal warping, despite being small in magnitude, has important effects on the low energy physics of few-layer graphene~\cite{Faraday_rotation_Morimoto2012,Trigonal_Koshino2009,Trigonal_Varlet2014}.  It stretches the Fermi circle along three directions in the Brillouin zone enforcing discrete three-fold rotational symmetry on the Fermi surface~\cite{Trigonal_Koshino2009}. The consequence of trigonal warping in bilayer graphene and in ABC-stacked TLG is the appearance of Lifshitz transition at low energy and the associated three-fold degenerate LLs which shows up as multiple of 3 in filling factors~\cite{Trigonal_Koshino2009,Trigonal_Varlet2014}. In ABA-stacked TLG, however, the effect of trigonal warping is more interesting as it introduces additional selection rules for the coupling between some LLs. The LLs cross each other, as a function of magnetic field (\textit{B}), \textit{E}$^\perp$ and filling factor. In presence of the electric field, coupling due to trigonal warping converts some of the level crossings into anticrossings~\cite{serbyn_new_2013,shimazaki_landau_2016} when the LL indices differ by 3. We use these LL crossings to determine the key band parameters.  An important aspect of the few-layer graphene physics, especially the electronic interaction, depends on quantitatively understanding the charge and electric field distribution. We exploit the extreme sensitivity of the LL crossing patterns to the non-uniformity in the \textit{E}$^\perp$ between the three layers of ABA-stacked TLG to present the first quantitative determination of these non-uniformities; this allows us to unequivocally infer the spin and valley resolved LL ordering. For the case of ABA-stacked TLG we find that the density (\textit{n}) of electrons  in the outer two layers is higher than the middle layer.

%------------------------------------------------------------------------------------------------------------------------
%Fig1
%------------------------------------------------------------------------------------------------------------------------

\begin{figure}[h]
\includegraphics[width=8.5cm]{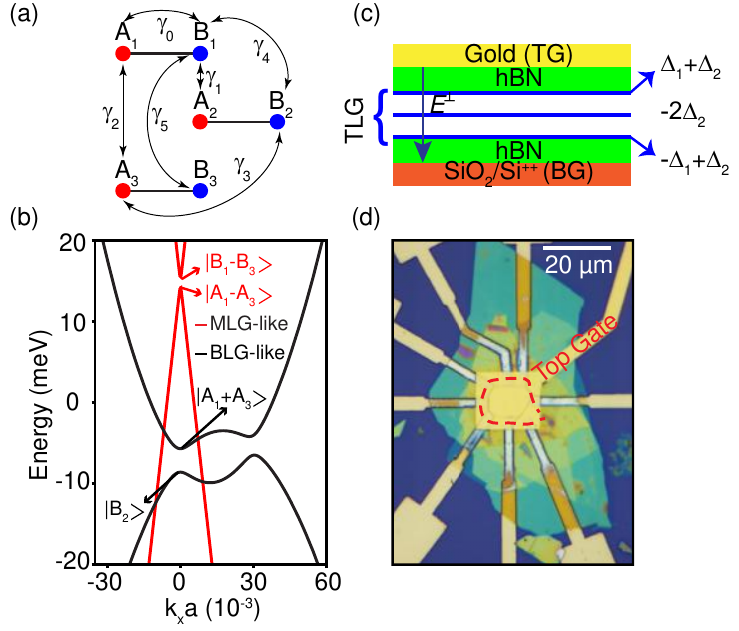}
\caption{ \label{fig:fig1} (a) A schematic of the ABA-stacked TLG unit cell showing the atomic sites along with all the hopping parameters. $\delta$ is the onsite energy difference between two inequivalent carbon atoms on the same layer. (b) Band structure of ABA-stacked TLG at zero electric field around \textit{K}$_+$ point ($\frac{4 \pi}{3}$,0) in the Brillouin zone. Red and black curves show MLG-like and BLG-like bands respectively. Wavefunctions of each band at (k$_x$,k$_y$)=(0,0) are polarized on different mirror symmetric and antisymmetric basis as labelled in the figure. (c) A schematic of the device showing the potential energy distribution across the three layers of the TLG in presence of an external  \textit{E}$^\perp$. Non-uniform charge distribution leads to a positive potential in the middle layer which decreases its electrostatic energy. (d) Optical image of the device on 300~nm SiO$_2$/Si$^{++}$ substrate after the top gate fabrication. Multiple hBN (greenish in colour) transferred during the fabrication are visible. The red dotted line shows the boundary of the TLG.}
\end{figure}

Fig.~\ref{fig:fig1}a and Fig.~\ref{fig:fig1}b show the unit cell of the ABA-stacked TLG and the calculated band structure using tight-binding formalism respectively (see Fig.S1 in Supplemental Material for band structure evolution with \textit{E}$^\perp$~\cite{footnote1}). Potential energy of each layer of the biased TLG~\cite{koshino_gate-induced_2009} is shown in the schematic Fig.~\ref{fig:fig1}c. The energy difference of the top and the bottom layer is 2$\Delta_1$. Theoretically, an average \textit{E}$^\perp$ inside the TLG can be defined as \textit{E}$^\perp_{\mathrm{av}}$=$\frac{2 \Delta_1}{(e)d}$ where d=0.67~nm is the separation of top and bottom layer of the TLG. If the charge distribution in three layers of ABA-stacked TLG is not same, it gives rise to a difference in the middle layer potential from the average of the top and bottom layers; this difference is defined to be proportional to a parameter $\Delta_2$. Non-zero $\Delta_2$ also generates a non-uniform electric field perpendicular to the layers. In ABA-stacked TLG low energy electronic states are centered on A$_1$-A$_3$, B$_1$-B$_3$ atomic sites (for MLG-like bands) and on A$_1$+A$_3$, B$_2$ atomic sites (for BLG-like bands)~\cite{serbyn_new_2013}. So, electrons primarily prefer to lie on the outer layer atoms to decrease the Coulomb energy by maximizing their spatial separation. This can lead to a positive potential (hence a positive $\Delta_2$) in the middle layer.

Fig.~\ref{fig:fig1}d shows an optical image of the dual-gated hexagonal boron nitride sandwiched ABA-stacked TLG device. The fabrication process is similar to the earlier published methods~\cite{dean_boron_2010,geim_van_2013,wang_one-dimensional_2013}. the mobility of our device is very high ($\sim$800,000~cm$^{2}$V$^{-1}$s$^{-1}$) which can be calculated from the field effect gating measurement (Fig.S2 in Supplemental Material~\cite{footnote1}). Low LL broadening $\sim$1~meV in our device (see Fig.S3 in Supplemental Material for determination of LL broadening~\cite{footnote1}) allows us to do quantum Hall (QH) measurements at very low \textit{B}. All the measurements have been done in a He-3 cryostat at 300~mK unless stated otherwise.

%------------------------------------------------------------------------------------------------------------------------
%Fig2
%------------------------------------------------------------------------------------------------------------------------
%

\begin{figure}[h]
\includegraphics[width=8.5cm]{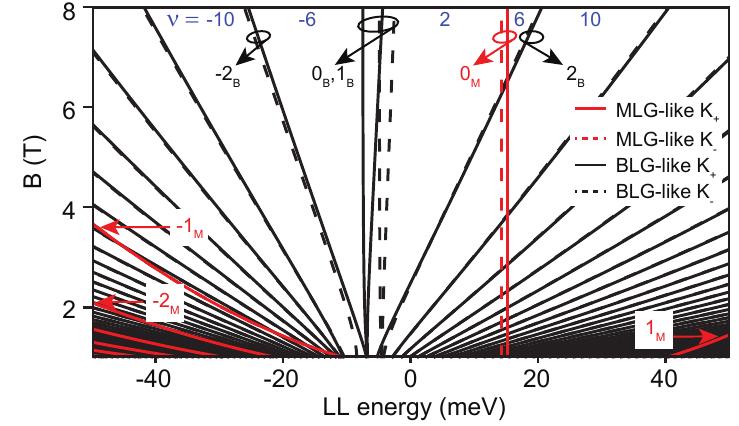}
\caption{ \label{fig:fig2} LL diagram as a function of magnetic field. Spin is kept degenerate in this diagram to avoid clutter. Red and black lines denote the MLG-like and BLG-like LLs respectively. Solid and dashed lines denote LLs from K$_+$ and K$_-$ valleys respectively.}
\end{figure}

In order to understand our experiments, we calculate LL spectrum of the system using the tight binding model (for details of the calculation see Supplemental Material~\Rmnum{9}~\cite{footnote1}). Fig.~\ref{fig:fig2} shows  that at zero electric field LLs of the ABA-stacked TLG can be classified into MLG-like and BLG-like LLs~\cite{serbyn_new_2013}. Each type of LLs can be labelled by a composite quantum number \textit{N}$_\mathrm{S}^{\mathrm{v}\sigma}$ consisting of an integer index \textit{N}, a subband index S (M,B) for MLG-like and BLG-like bands, a valley index v (+,-) for \textit{K}$_+$ and \textit{K}$_-$ valleys and a spin index $\sigma$ ($\uparrow$,$\downarrow$) for up and down spins respectively. LLs without spin or valley index implies that spin or valley is degenerate. LLs from the MLG-like bands disperse with magnetic field as $\sim\sqrt{\textit{B}}$ whereas LLs from BLG-like bands disperse as $\sim$\textit{B}.

%------------------------------------------------------------------------------------------------------------------------
%Fig3
%------------------------------------------------------------------------------------------------------------------------
%

\begin{figure}[h]
\includegraphics[width=8.5cm]{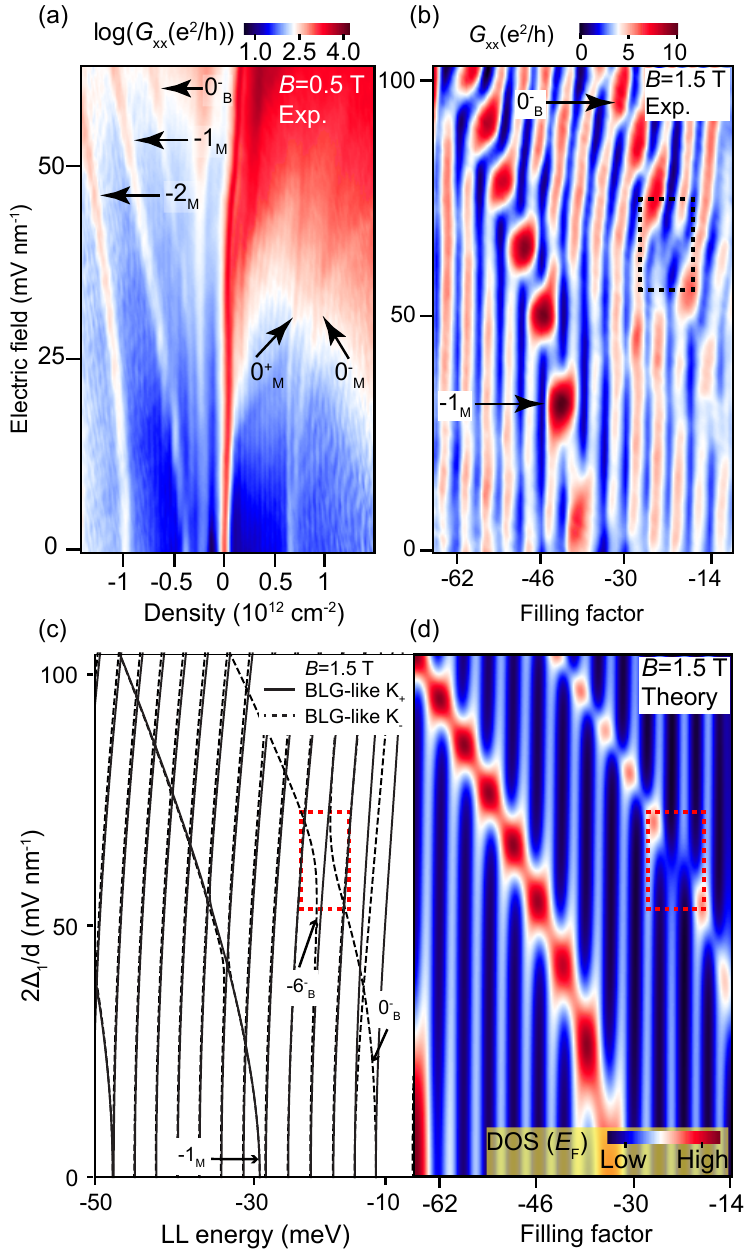}
\caption{ \label{fig:fig3} (a) and (b) Colour plots of experimental $G_{\mathrm{xx}}$ at 0.5~T and 1.5~T respectively. The thick curved  lines are the locus of the LL crossings. Here, 0$^-_\mathrm{B}$ refers to the locus of all the crossings between 0$^-_\mathrm{B}$ and other BLG-like LLs. Other labels have a similar meaning. On the hole side of (a) the first thick curve from the right (near zero density) comes due to a group of closely spaced LL crossing with each other. It does not have a particular LL index and hence it is not labelled. (c) LL energy diagram at 1.5~T which shows 0$^-_\mathrm{B}$ and -1$_\mathrm{M}$ LLs rapidly dispersing with electric field crossing several BLG-like LLs. The red dashed rectangle shows the anticrossing between 0$^-_\mathrm{B}$ and -6$^-_\mathrm{B}$ LLs. (d) Calculated DOS(\textit{E}$_\mathrm{F}$) from the LLs shown in (c).}
\end{figure}

We study the \textit{E}$^\perp$ dependence of longitudinal conductance (\textit{G}$_{\mathrm{xx}}$) for different values of magnetic field and encounter LL crossings or anticrossings depending on the symmetries of the underlying LLs. Fig.~\ref{fig:fig3}a and Fig.~\ref{fig:fig3}b show  \textit{G}$_{\mathrm{xx}}$ plots as a function of \textit{E}$^\perp$ and  density  at 0.5~T and 1.5~T respectively. At such low \textit{B}, LLs are not fully resolved and hence locus of the LL crossing points form curves in \textit{n}-\textit{E}$^\perp$ space.  On the electron side at 0.5~T, Fig.~\ref{fig:fig3}a shows two thick curves (labelled 0$_\mathrm{M}^+$ and 0$_\mathrm{M}^-$) whose separation increases in density axis with increasing \textit{E}$^\perp$. On the hole side we see four thick curves dispersing with \textit{E}$^\perp$. At a larger \textit{B} of 1.5~T (Fig.~\ref{fig:fig3}b), we resolve the LLs more clearly and see two thick curves on the hole side. The first curve from the right labelled 0$_\mathrm{B}^-$ shows an interesting splitting pattern at \textit{E}$^\perp$$\sim$65 mVnm$^{-1}$  between $\nu$=-22 and -26 marked by a dashed rectangle.

We determine the band parameters by matching the experimental LL crossings with theory at 1.5~T. We find $\gamma_0$=3.1~eV, $\gamma_1$=390~meV, $\gamma_2$=-20~meV, $\gamma_3$=315~meV, $\gamma_4$=120~meV,  $\gamma_5$=18~meV,  $\delta$=20~meV  best describe our data. Determination of $\Delta_2$ requires input from high \textit{B} data and is discussed later (Details of fitting are described in Supplemental Material~\Rmnum{10}~\cite{footnote1}). Fig.~\ref{fig:fig3}c and Fig.~\ref{fig:fig3}d are the numerically calculated LL spectrum as a function of \textit{E}$^\perp$ at 1.5~T and the corresponding density of states at Fermi energy (DOS(\textit{E}$_\mathrm{F}$)) which matches well with our experimental data.

Now, using  the experimental data as shown in Fig.~\ref{fig:fig3}a on the electron side we can estimate the band gap between the MLG-like bands. We note that the 0$_\mathrm{M}^+$ and 0$_\mathrm{M}^-$ marked LLs in Fig.~\ref{fig:fig3}a originate from the conduction and valence band edges of MLG-like bands as the bands are split in ABA-stacked TLG~\cite{serbyn_new_2013}. So, the valley gap of 0$_\mathrm{M}$ LL is identical to the band gap of MLG-like bands. From our experiment, we see that the valley splitting of the 0$_\mathrm{M}$ LL cannot be resolved at zero electric field (Fig.~\ref{fig:fig3}a) meaning the valley splitting is $\sim$1~meV for a LL broadening of $\sim$1~meV; this means that the band gap of MLG-like bands is also $\sim$1~meV. Calculation  shows  that the valley gap of 0$_\mathrm{M}$ LL starts increasing from 1~meV with increasing \textit{E}$^\perp$ (see Fig.S7 and Fig.S12 in Supplemental Material for the LL diagram and the calculated DOS respectively~\cite{footnote1}). This band gap was assumed to be much larger in many previous studies~\cite{taychatanapat_quantum_2011,stepanov_tunable_2016,shimazaki_landau_2016} (Table~\Rmnum{1} in Supplemental Material~\cite{footnote1}).

Now we focus on the hole side. The second curve from the right labelled 0$_\mathrm{B}^-$ in  Fig.~\ref{fig:fig3}a is the locus of the crossings of 0$_\mathrm{B}^-$ LL with other BLG-like LLs. The successive curves on the hole side appear from the crossings of MLG-like LLs with other BLG-like LLs i.e. the third curve is due to -1$_\mathrm{M}$, the fourth curve is due to -2$_\mathrm{M}$ and so on. Similarly, at 1.5~T these LLs are labelled in Fig.~\ref{fig:fig3}b. Now we try to understand the feature marked by a dashed rectangle in  Fig.~\ref{fig:fig3}b which is an anticrossing. This anticrossing appears due to the trigonal warping ($\gamma_3$) which couples a BLG-like LL with every third other BLG-like LLs~\cite{koshino_landau_2011,serbyn_new_2013}. This can be understood physically in the following manner. Low energy BLG-like LLs lies predominantly on the A$_1$+A$_3$ and B$_2$ lattice sites which get coupled in the presence of $\gamma_3$, as $\gamma_3$ is the hopping parameter between A$_{1,3}$ $\leftrightarrow$ B$_2$ sites. Calculated LL spectra at 1.5~T (Fig.~\ref{fig:fig3}c) shows that the feature at \textit{E}$^\perp$$\sim$65~mVnm$^{-1}$ in Fig.~\ref{fig:fig3}b corresponds to the intra-subband anticrossing between 0$_\mathrm{B}^-$  and -6$_\mathrm{B}^-$ LLs (for more details see Supplemental Material~\Rmnum{4}~\cite{footnote1}). A closer look at the 0.5~T data at \textit{n}$\sim$-0.5$\times$10$^{12}$cm$^{-2}$ (Fig.~\ref{fig:fig3}a) shows a vertical line like feature joining second (marked 0$^-_\mathrm{B}$) and third (marked -1$_\mathrm{M}$) curves on the hole side which is also a signature of anticrossing (see Fig.S6d and Fig.S8 in Supplemental Material for the DOS and the LL diagram~\cite{footnote1}). We find good agreement between theoretical and experimental \textit{E}$^\perp$ values at the LL crossing points which shows that the change of screening with \textit{E}$^\perp$ or \textit{n} does not play a major role in ABA-stacked TLG at low \textit{B}.

%------------------------------------------------------------------------------------------------------------------------
%Fig4
%------------------------------------------------------------------------------------------------------------------------

\begin{figure}[h]
\includegraphics[width=8.5cm]{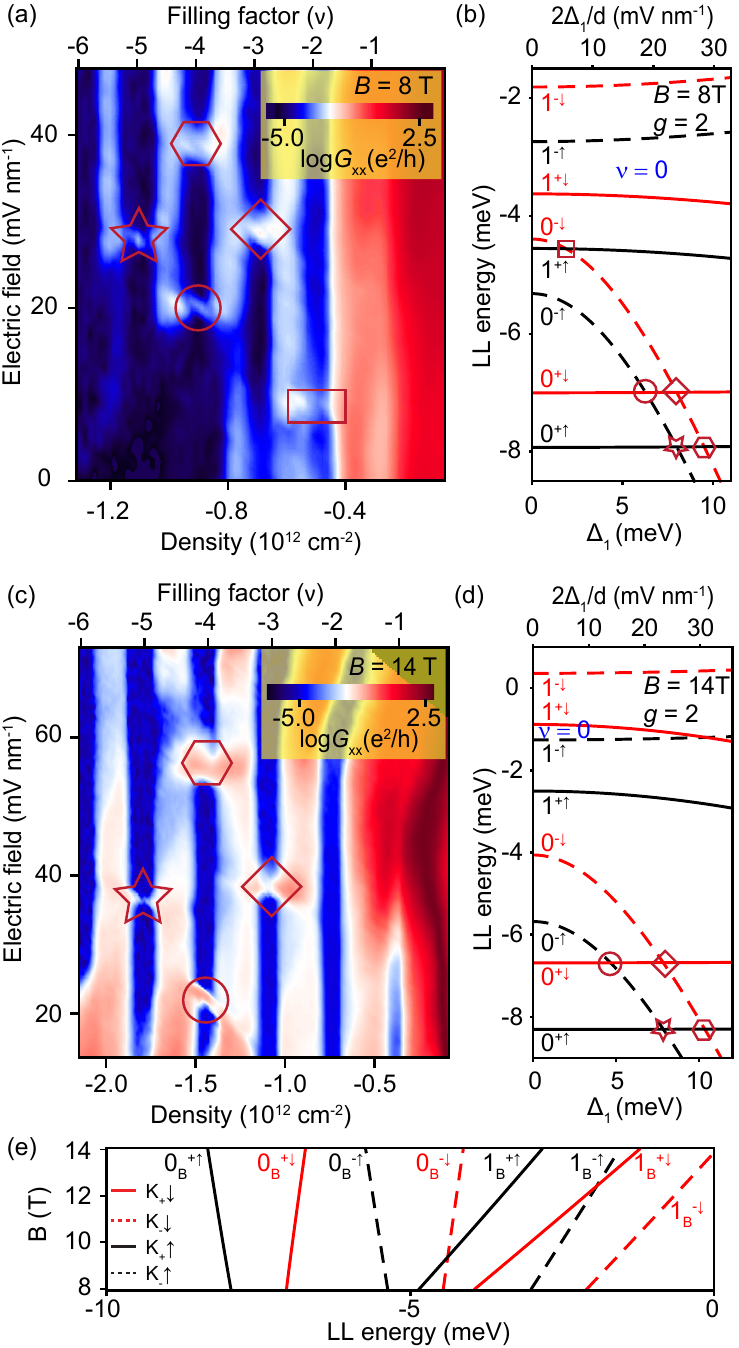}
\caption{\label{fig:fig4} (a) and (c) Experimental \textit{G}$_{\mathrm{xx}}$  showing the transitions between different QH states as a function of \textit{E}$^\perp$ at 8~T and 14~T respectively. (b) and (d) Calculated LL spectra at 8~T and 14~T showing the BLG-like LLs corresponding to the experimental data. Red and black lines denote the LLs with $\downarrow$ and $\uparrow$ spins respectively. Solid and dashed lines denote the LLs from \textit{K}$_+$ and \textit{K}$_-$ valleys respectively. $\square$, $\Diamond$, $\star$ symbols designate LL crossings at $\nu$=-2,-3 and -5. $\bigcirc$ and  $\hexagon$ symbols denote the  lower and higher electric field crossings  respectively at $\nu$=-4. (e) Dispersion of the LLs with magnetic field which shows the change in 0$_\mathrm{B}^{-\downarrow}$ and 1$_\mathrm{B}^{+\uparrow}$ LL ordering in going from 8~T to 14~T.}
\end{figure}

We now show that \textit{E}$^\perp$ inside ABA-stacked TLG can be very non-uniform, especially at low energies (for zeroth LLs) and at small $\Delta_1$ because the magnitude of  $\Delta_2$ is comparable to the separation between this manifold. A simple estimation of this non-uniformity is shown in Supplemental Material~\Rmnum{7}~\cite{footnote1}. For this reason, we determine $\Delta_2$ from the crossing within zeroth LLs at high \textit{B} (Fig.~\ref{fig:fig4}) for the first time. Controlling the top and back gate voltages independently we study the LL crossings as a function of \textit{E}$^\perp$. At 8~T, there is one LL crossing at each filling factor $\nu$=-2,-3 and -5, but  at $\nu$=-4 there are two LL crossings, as seen in Fig.~\ref{fig:fig4}a. Increasing \textit{B} from 8~T to 14~T leads to the disappearance of the crossing at $\nu$=-2 (Fig.~\ref{fig:fig4}c) changing the crossing pattern. From our calculated LL spectra at high magnetic fields (Fig.~\ref{fig:fig4}b and Fig.~\ref{fig:fig4}d), we find that the crossing pattern of the low energy LLs is very sensitive to the value of $\Delta_2$~\cite{serbyn_new_2013}. Hence we use this change of crossing pattern to determine $\Delta_2$ by matching the patterns simultaneously at 8~T and 14~T. We note that the valley splitting of 0$_\mathrm{B}$ LL at $\Delta_1$=0 is \textit{E}$_{\mathrm{B}}^{0+}$-\textit{E}$_{\mathrm{B}}^{0-}$=$\frac{|\gamma_2|}{2}$-3$\Delta_2$ which is very sensitive to $\Delta_2$ as it comes with a factor of 3. So, increasing $\Delta_2$ from zero decreases the valley splitting and their order eventually flips for $\Delta_2>\frac{|\gamma_2|}{6}$=3.3~meV, resulting in the 0$_\mathrm{B}^-$ LL lying above 0$_\mathrm{B}^+$ LL~\cite{footnote2}. This is a crucial fact because depending on the order of the LLs the crossing pattern can change completely.  Matching the experimental crossing pattern with theory we realised that the value of $\Delta_2$ has to be such that 0$_\mathrm{B}^{-\downarrow}$ lies between spin split 1$_\mathrm{B}^{+\uparrow}$  and 1$_\mathrm{B}^{+\downarrow}$ at 8~T for $\Delta_1$=0. We find a narrow range of positive $\Delta_2\sim$4.3-4.4~meV which can explain all our experimental data. We consider $\Delta_2$=+4.3~meV in all the calculations; which suggests that the weight of the electronic wavefunction is concentrated more on the outer two layers. The existence of a positive potential in the middle layer can also be understood in QH regime~\cite{serbyn_new_2013}  by noticing that $\nu$=-5 to -2 range on the hole side corresponds to emptying out either 1$_\mathrm{B}^+$ or both 1$_\mathrm{B}^+$ and 0$_\mathrm{B}^+$ LLs (Fig.~\ref{fig:fig4}b and Fig.~\ref{fig:fig4}d). Both of these states are polarized in the middle layer, emptying out these states can create a positive potential in the middle layer~\cite{serbyn_new_2013}.

\clearpage

Analyzing the LL crossing patterns at 8~T and 14~T we can uniquely determine the ordering of spin and valley resolved LLs. Hence, we can label the LLs responsible for the observed crossings which was not possible in the previous studies~\cite{campos_landau_2016,shimazaki_landau_2016}. Fig.~\ref{fig:fig4}b and Fig.~\ref{fig:fig4}d show the calculated LL spectra  on the hole side at 8~T and 14~T respectively. We note that the key change in the LL diagram in going from 8~T to 14~T is flipping the order of 0$_\mathrm{B}^{-\downarrow}$ and 1$_\mathrm{B}^{+\uparrow}$ LLs which is shown in Fig.~\ref{fig:fig4}e. This happens because orbital energy of 1$_\mathrm{B}^{+\uparrow}$ LL increases linearly with \textit{B} whereas energy of 0$_\mathrm{B}^{-\downarrow}$ LL is independent of \textit{B} other than a small change due to Zeeman interaction. Calculation suggests that the energy of the 0$_\mathrm{B}^-$  LL decreases rapidly with increasing  $\Delta_1$ leading to the crossings with other LLs below it. So, the crossing pattern depends on the sequence of the LLs below 0$_\mathrm{B}^-$ LL.  In our experimental range of electric field, 0$_\mathrm{B}^-$, 0$_\mathrm{B}^+$ and 1$_\mathrm{B}^+$ LLs are solely responsible for all the crossings between $\nu$ =-5 and -2. LL diagrams shown in Fig.~\ref{fig:fig4}b and Fig.~\ref{fig:fig4}d explain the LL crossings observed in experiment (Fig.~\ref{fig:fig4}a and Fig.~\ref{fig:fig4}c).

In summary, we have studied the role of the electric field in the LL crossing physics in ABA-stacked TLG. At low \textit{B} we have shown that the electric field allows us to observe from simple LL crossings to anticrossings mediated by trigonal warping which we find in the experiment as an additional selection rule for the coupling between two LLs. Matching the LL crossing pattern at multiple magnetic fields allows us to pin down $\Delta_2$$\sim$4.3~meV which in turn helps us to determine the order of the low energy LLs uniquely. This value  is surprisingly close to the self-consistently calculated value of $\Delta_2$$\sim$4~meV~\cite{serbyn_new_2013}. The electronic interaction in few-layer graphene~\cite{jang_stacking_2015,Apalkov_2012,dean_multicomponent_2011,maher2014tunable} depends quantitatively on understanding the charge and electric field distribution in which our work provides an important direction. Recently predicted~\cite{Jain_graphene} electric field tunable non-Abelian fractional quantum Hall states indicates the possibility to realize these exotic states in trilayer graphene.

We thank Sameer Grover, Sreejith GJ, Vibhor Singh, Fengcheng Wu and Sajal Dhara for helpful discussions. Biswajit Datta is a recipient of Prime Minister’s Fellowship Scheme for Doctoral Research, a public-private partnership between Science \& Engineering Research Board (SERB), Department of Science \& Technology, Government of India and Confederation of Indian Industry (CII). His host institute for research is Tata Institute of Fundamental Research, Mumbai and the partner company is Tata Steel Ltd. We acknowledge Swarnajayanti Fellowship of Department of Science and Technology (for MMD), Nanomission grant SR/NM/NS-45/2016 and Department of Atomic Energy of Government of India for support. Preparation of hBN single crystals are supported by the Elemental Strategy Initiative conducted by the MEXT, Japan and JSPS KAKENHI Grant Number JP15K21722.

%

%-----------------------------------------------------------------------------------------------------------------------------------------------
%                                                             Supplementary Material
%-----------------------------------------------------------------------------------------------------------------------------------------------

%%%%%%%%%% Merge with supplemental materials %%%%%%%%%%
\pagebreak
\widetext
\begin{center}
\textbf{\large Supplemental Material: Landau level diagram and the continuous rotational symmetry breaking in trilayer graphene}
\end{center}
%%%%%%%%%% Merge with supplemental materials %%%%%%%%%%
%%%%%%%%%% Prefix a "S" to all equations, figures, tables and reset the counter %%%%%%%%%%

\setcounter{equation}{0}
\setcounter{figure}{0}
\setcounter{table}{0}
\setcounter{page}{1}

\renewcommand{\theequation}{S\arabic{equation}}
\renewcommand{\thefigure}{S\arabic{figure}}
\renewcommand{\thepage}{S\arabic{page}}
\renewcommand{\bibnumfmt}[1]{[S#1]}
\renewcommand{\citenumfont}[1]{S#1}

%\onecolumngrid
%\makeatletter

%%%%%%%%%% Prefix a "S" to all equations, figures, tables and reset the counter %%%%%%%%%%

\section{Band-structure evolution of ABA-stacked trilayer graphene with electric field}

\begin{figure}[h]
\includegraphics[width=15.5cm]{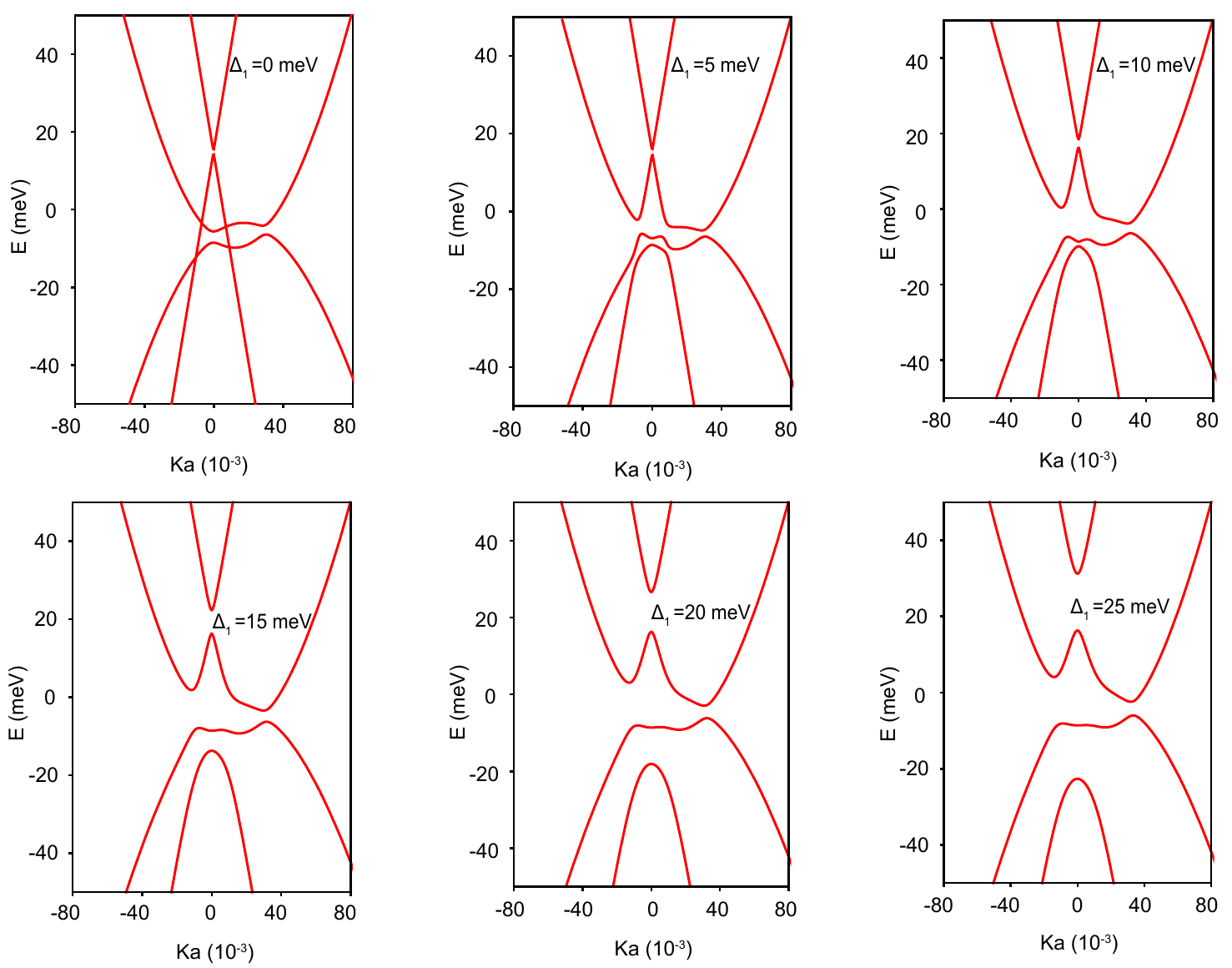}
\caption{ \label{fig:Sfig1} Band structure of ABA-stacked trilayer graphene along k$_\mathrm{x}$ direction around \textit{K}$_\mathrm{+}$ valley.}
\end{figure}

Fig.~\ref{fig:Sfig1}  shows the calculated band structure evolution  of ABA-stacked TLG along k$_x$  direction with electric field. All the tight-binding parameters used are mentioned in the main text. It is clear that increasing electric field  hybridizes the bands. Hybridization also changes the band gap between the MLG-like bands. It is clear that for small electric fields (up to $\Delta_1$=5~meV in the figure) the band gap between the MLG-like bands decreases with increasing electric field. Increasing electric field beyond that leads to the enhancement of the band gap. In QH regime 0$_\mathrm{M}^+$ and 0$_\mathrm{M}^-$ LLs originate from the bottom of the MLG-like conduction band and the top of the MLG-like valence bands respectively. So, the valley gap between these LLs is also the band gap of MLG-like bands at zero magnetic field and hence follows the same electric field dependence.

\clearpage

\section{Field effect gating measurement}
\begin{figure}[h]
\includegraphics[width=15.5cm]{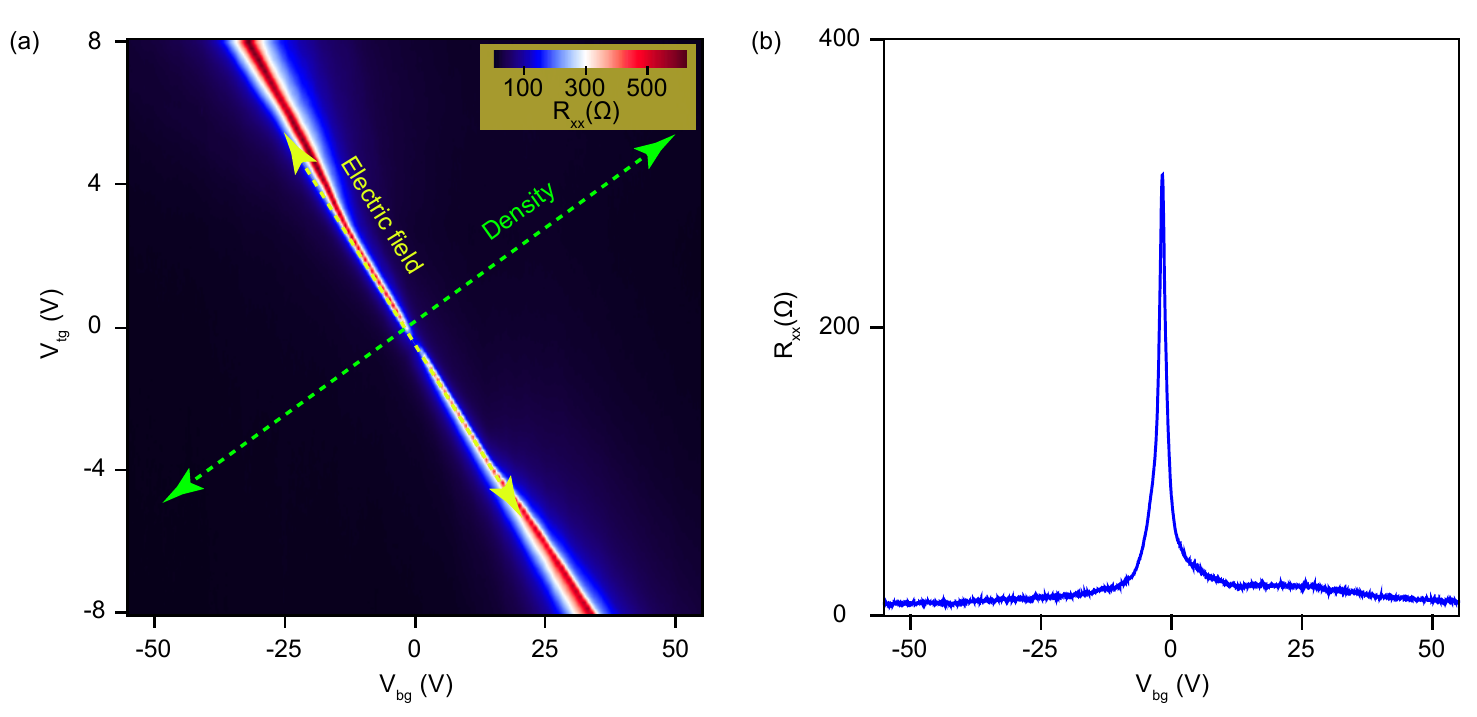}
\caption{ \label{fig:Sfig2} (a) Colour plot of four-probe resistance at 0.3~K. (b) Line slice of the colour plot at \textit{V}$_{\mathrm{tg}}$=0~V.}
\end{figure}

Fig.~\ref{fig:Sfig2}a shows the colour plot of four-probe resistance as a function of top-gate and back-gate voltages. The yellow and green arrows designate the direction of electric field and density respectively. \textit{E}$^\perp$ and \textit{n} can be estimated from the experiment by
\textit{E}$^\perp$=\textit{E}$^\perp_\mathrm{bg}$+\textit{E}$^\perp_\mathrm{tg}$=$\frac{\textit{C}_\mathrm{bg}\textit{V}_\mathrm{bg}}{2\kappa_\mathrm{bg}\epsilon_\mathrm{0}}$-$\frac{\textit{C}_\mathrm{tg}\textit{V}_\mathrm{tg}}{2\kappa_\mathrm{tg}\epsilon_\mathrm{0}}$ and \textit{n}=$\frac{1}{e}[\textit{C}_\mathrm{bg}\textit{V}_\mathrm{bg}+\textit{C}_\mathrm{tg}\textit{V}_\mathrm{tg}]$ where \textit{V}$_\mathrm{bg}$ (\textit{V}$_\mathrm{tg}$) is the back (top)-gate voltage, \textit{C}$_\mathrm{bg}$ (\textit{C}$_\mathrm{tg}$) is the capacitance per unit area of the back (top)-gate, $\kappa_\mathrm{bg}$ ($\kappa_\mathrm{tg}$) is the dielectric constant of back (top)-gate dielectric and $\epsilon_\mathrm{0}$ is the vacuum permittivity. \textit{C}$_\mathrm{bg}$ and \textit{C}$_\mathrm{tg}$ are extracted from the QH data: \textit{C}$_\mathrm{bg}$=104~\textmu Fm$^{-2}$ and $\frac{\textit{C}_\mathrm{tg}}{\textit{C}_\mathrm{bg}}$=4. The sharp gating curve (Fig.~\ref{fig:Sfig2}b) is a typical for high mobility devices. The mobility of this device at 1.5~K is $\sim$800,000~cm$^{2}$V$^{-1}$s$^{-1}$.

\clearpage

\section{Quantum mobility from SdH oscillations}
\begin{figure}[h]
\includegraphics[width=15.5cm]{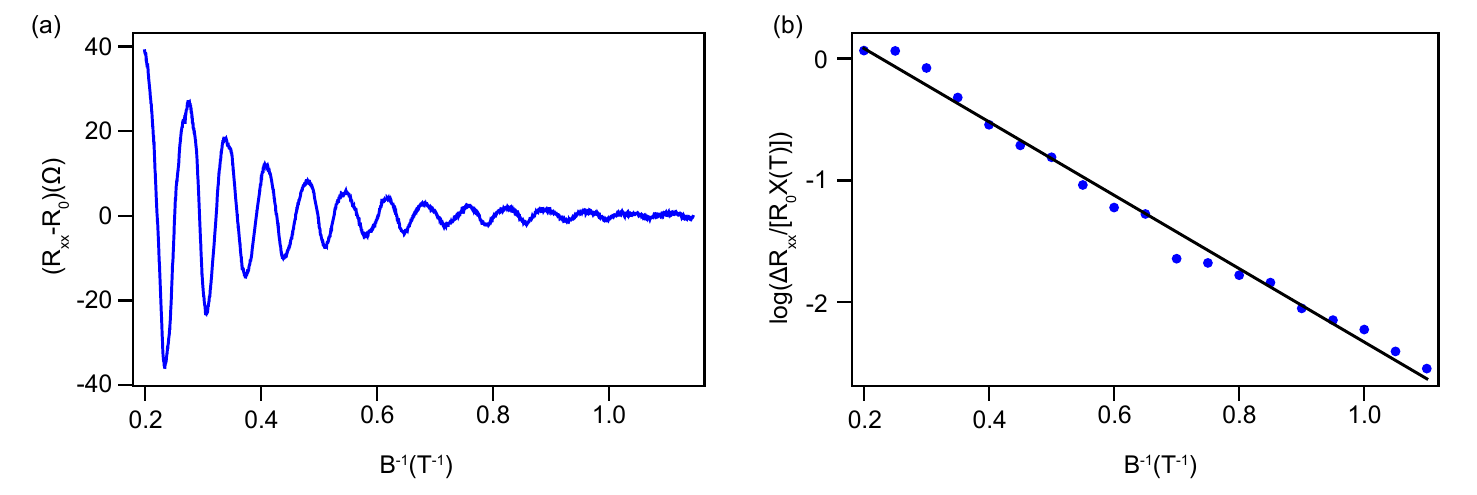}
\caption{ \label{fig:Sfig3} (a) SdH oscillations after subtracting the non-oscillatory background. (b) Straight-line fit in the Dingle plot. }
\end{figure}

We estimate disorder strength ($\Gamma$) and quantum scattering time ($\tau_\mathrm{q}$) from the magnetic field dependence of the SdH oscillations~\cite{coleridge_small-angle_1991_s,Das_Sarma_transport_scatter_2008_s,hong2009quantum_s}. Single particle quantum broadening ($\Gamma$) of the  LLs  is related to the quantum scattering time ($\tau_\mathrm{q}$) by $\Gamma = \frac{\hbar}{2\tau_\mathrm{q}}$. Amplitude of the SdH oscillations is given by~\cite{coleridge_small-angle_1991_s,Das_Sarma_transport_scatter_2008_s} $\frac{\Delta R}{R_\mathrm{0}} = 4 X(T) exp(-\pi/\omega_\mathrm{c} \tau_\mathrm{q})$ where $\omega_\mathrm{c} = \frac{e B}{m^*}$ is the cyclotron frequency. $\Delta R$ and $R_\mathrm{0}$  are the oscillatory and non-oscillatory part of the resistance respectively. $X(T)$ is the temperature dependent amplitude $X(T) = \frac{2 \pi^2 k_\mathrm{B} T / \hbar \omega_\mathrm{c}}{sinh(2 \pi^2 k_\mathrm{B} T / \hbar \omega_\mathrm{c})}$. Fig.~\ref{fig:Sfig3}a shows SdH oscillations at density 1.5$\times$10$^{12}$~cm$^{-2}$ after subtracting the non-oscillatory background resistance. Fig.~\ref{fig:Sfig3}b  shows the Dingle plot. Slope of the fitted line is given by $-\pi m^*/e \tau_\mathrm{q}$ which gives the quantum scattering time $\tau_\mathrm{q} \approx$ 287~fs and disorder energy $\Gamma \sim$1.1~meV. The quantum scattering time can be related to the quantum mobility $\mu_\mathrm{q}$=$\frac{e \tau_\mathrm{q}}{m^*}\approx$10,000~cm$^{2}$V$^{-1}$s$^{-1}$. SdH oscillations should be visible at a magnetic field B$_\mathrm{c}$ for which $\mu_\mathrm{q}$B$_\mathrm{c}\approx$~1; this gives B$_\mathrm{c}\approx$~1~T. This is consistent with our observation and affirms that appearance of SdH oscillation depends on quantum mobility instead of the transport mobility. The large ratio of transport scattering time and quantum scattering time in our device ($\frac{\tau_\mathrm{t}}{\tau_\mathrm{q}} \approx$ 50-80) is consistent with high mobility  two-dimensional electron systems~\cite{coleridge_small-angle_1991_s}.  Large $\tau_\mathrm{t}/\tau_\mathrm{q}$ indicates that small angle long range Coulomb scattering is the dominant scattering mechanism in our device~\cite{Das_Sarma_transport_scatter_2008_s,coleridge_small-angle_1991_s,Knap_interaction_in_SdH_s}.

\clearpage

\section{LL evolution with electric field at low magnetic field}

\begin{figure}[h]
\includegraphics[width=15.5cm]{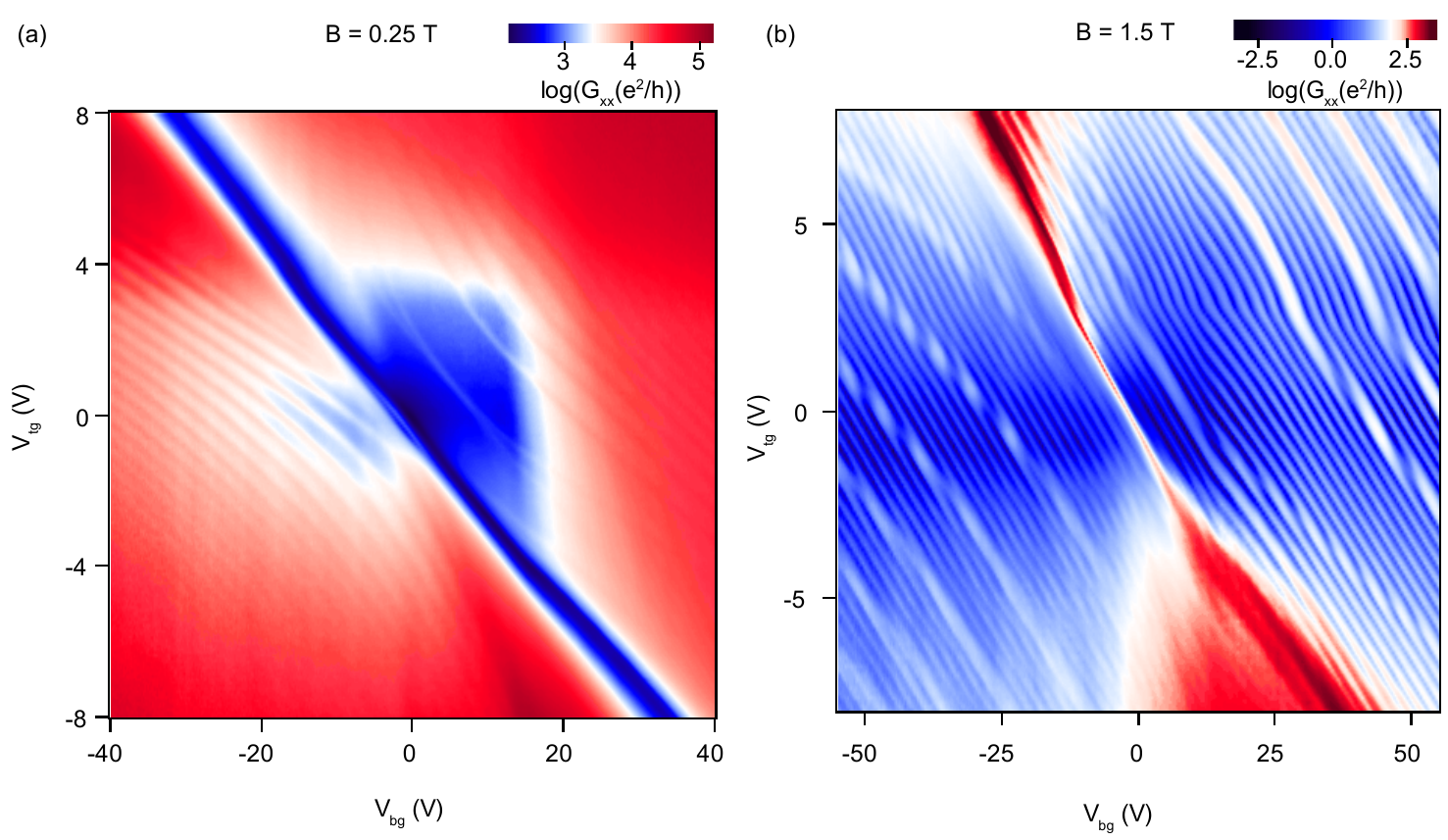}
\caption{ \label{fig:Sfig4} \textit{G}$_{\mathrm{xx}}$ as a function of \textit{V}$_{\mathrm{tg}}$ and \textit{V}$_{\mathrm{bg}}$ at (a) \textit{B}=0.25~T and (b) \textit{B}=1.5~T.}
\end{figure}

\begin{figure}[h]
\includegraphics[width=16cm]{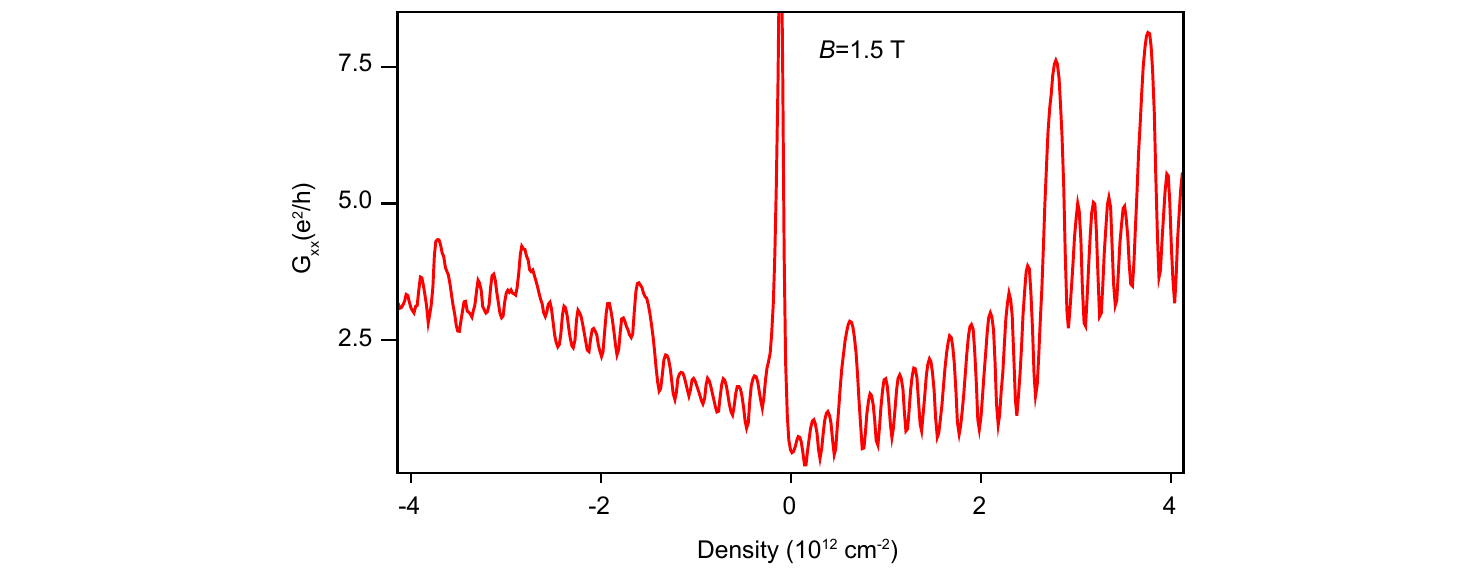}
\caption{ \label{fig:Sfig5}  \textit{G}$_{\mathrm{xx}}$ as a function of density at zero electric field.}
\end{figure}

\begin{figure}[h]
\includegraphics[width=16cm]{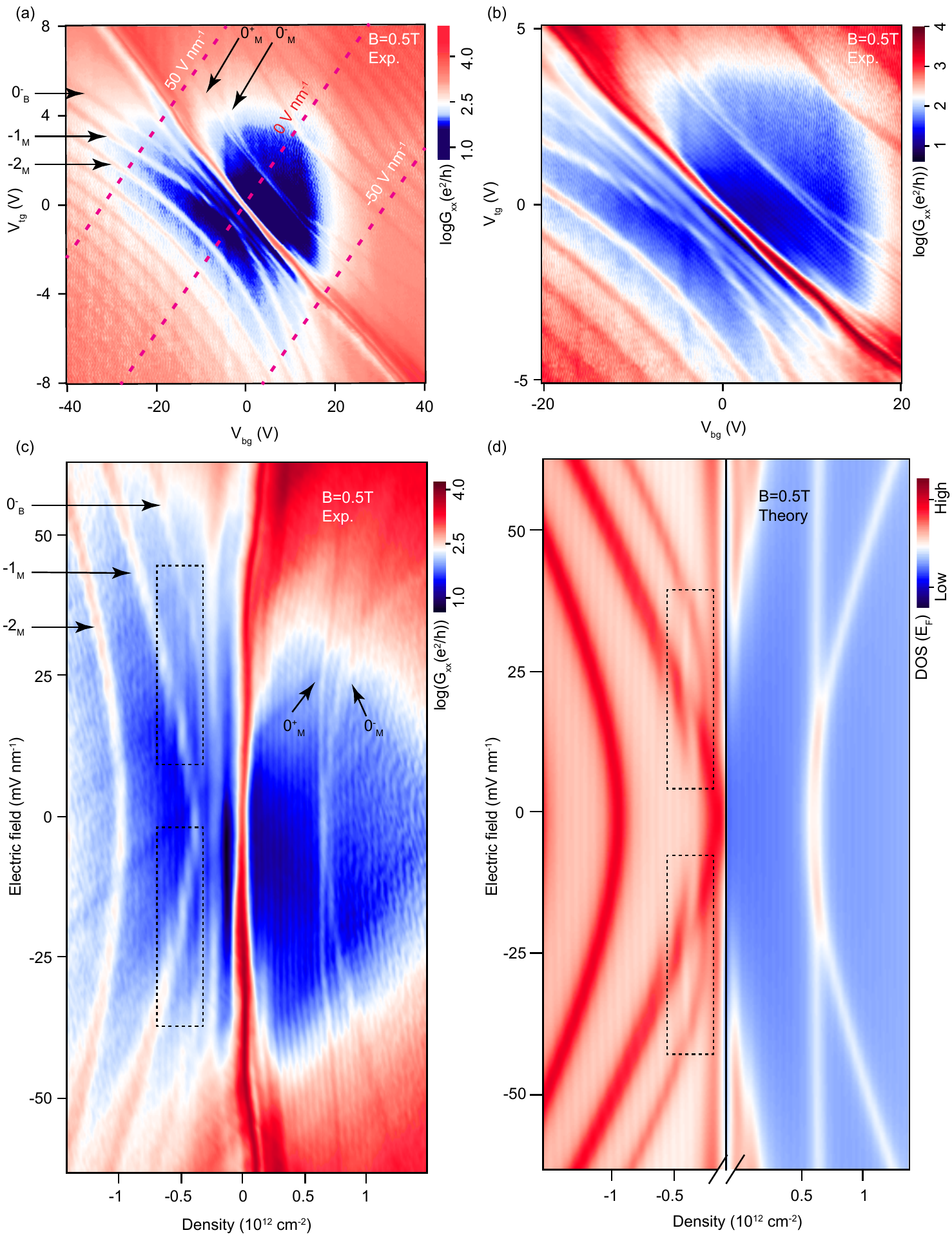}
\caption{ \label{fig:Sfig6} (a) Experimental colour plot of \textit{G}$_{\mathrm{xx}}$ as a function of \textit{V}$_{\mathrm{bg}}$ and \textit{V}$_{\mathrm{tg}}$ at 0.5~T. (b) Zoomed in plot of panel-a. (c) \textit{G}$_{\mathrm{xx}}$ plotted as a function of electric field and density. (d) Theoretically calculated DOS(\textit{E}$_{\mathrm{F}}$).}
\end{figure}

\begin{figure}[h]
\includegraphics[width=15.5cm]{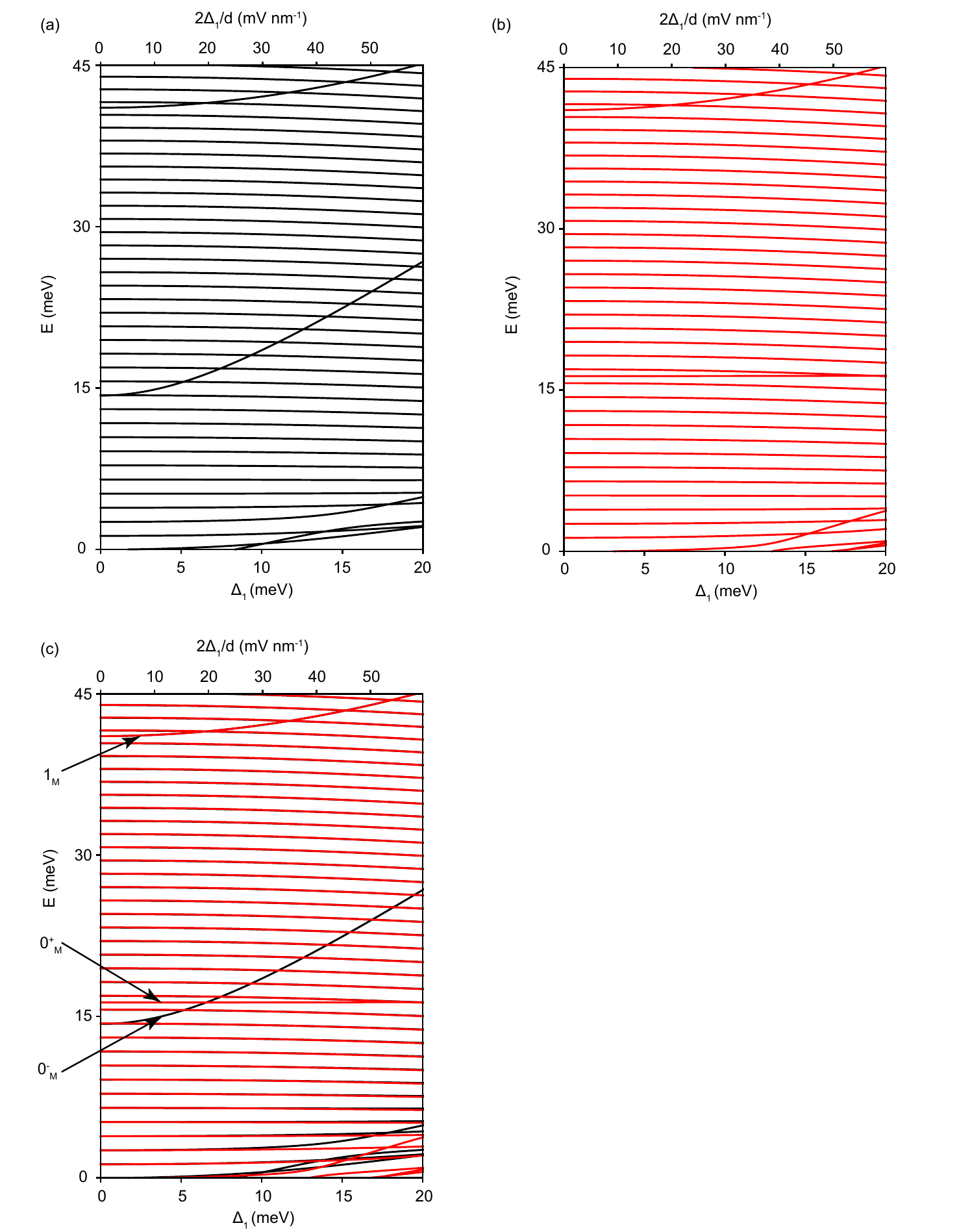}
\caption{ \label{fig:Sfig7} LL energy as a function of electric field at 0.5~T on the electron side (a) at 		\textit{K}$_-$ valley (b) at 		\textit{K}$_+$ valley (c) at both the valleys together.}
\end{figure}

\begin{figure}[h]
\includegraphics[width=15.5cm]{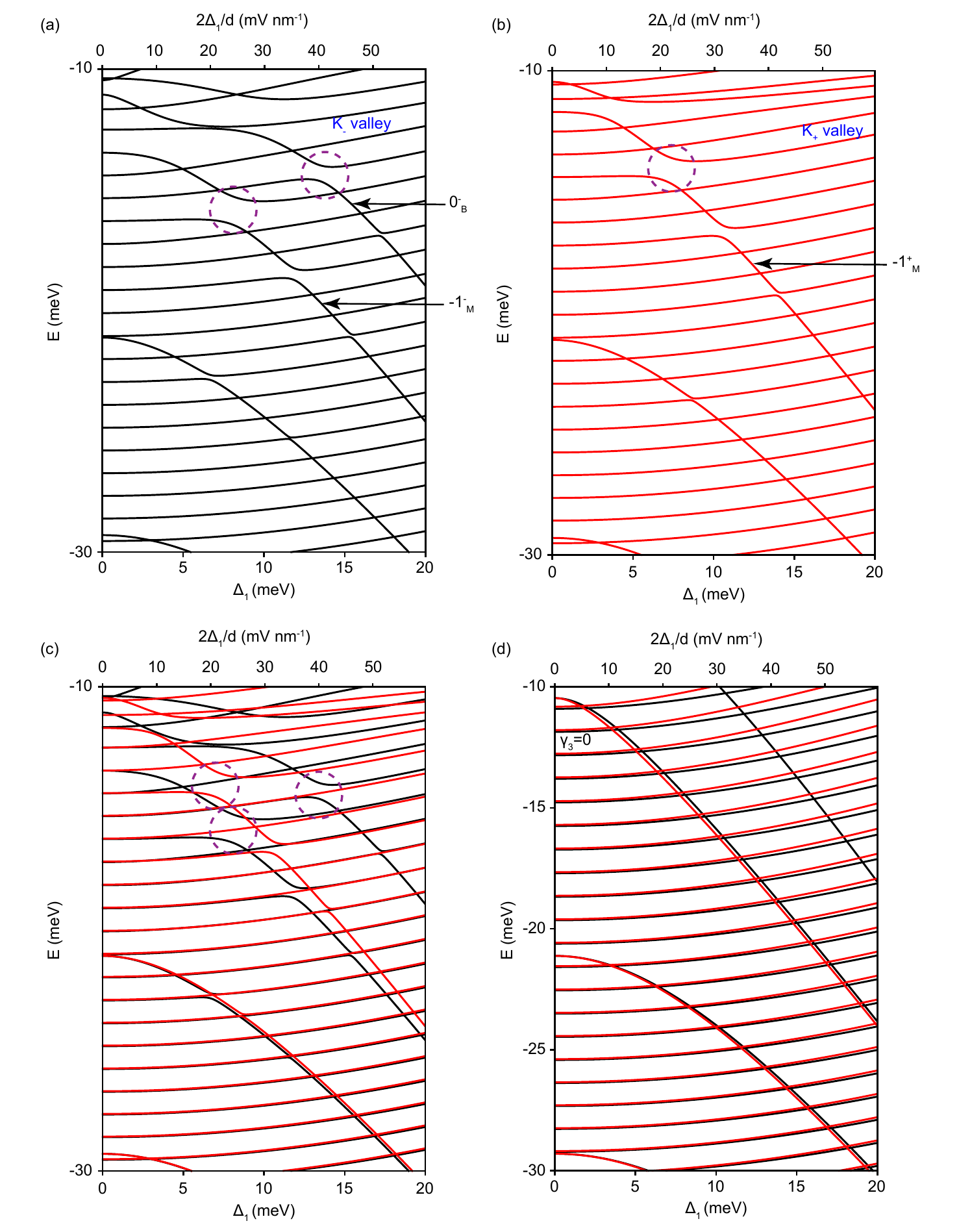}
\caption{ \label{fig:Sfig8} LL energy as a function of electric field at 0.5~T on the hole side showing the anticrossings between different LLs (a) at 		\textit{K}$_-$ valley (b) at 		\textit{K}$_+$ valley (c) at both the valleys together. (d) Calculated LL diagram setting $\gamma_3$=0 showing no anticrossings. This shows that the anticrossings are controlled by the trigonal warping of the band.}
\end{figure}

\begin{figure}[h]
\includegraphics[width=16cm]{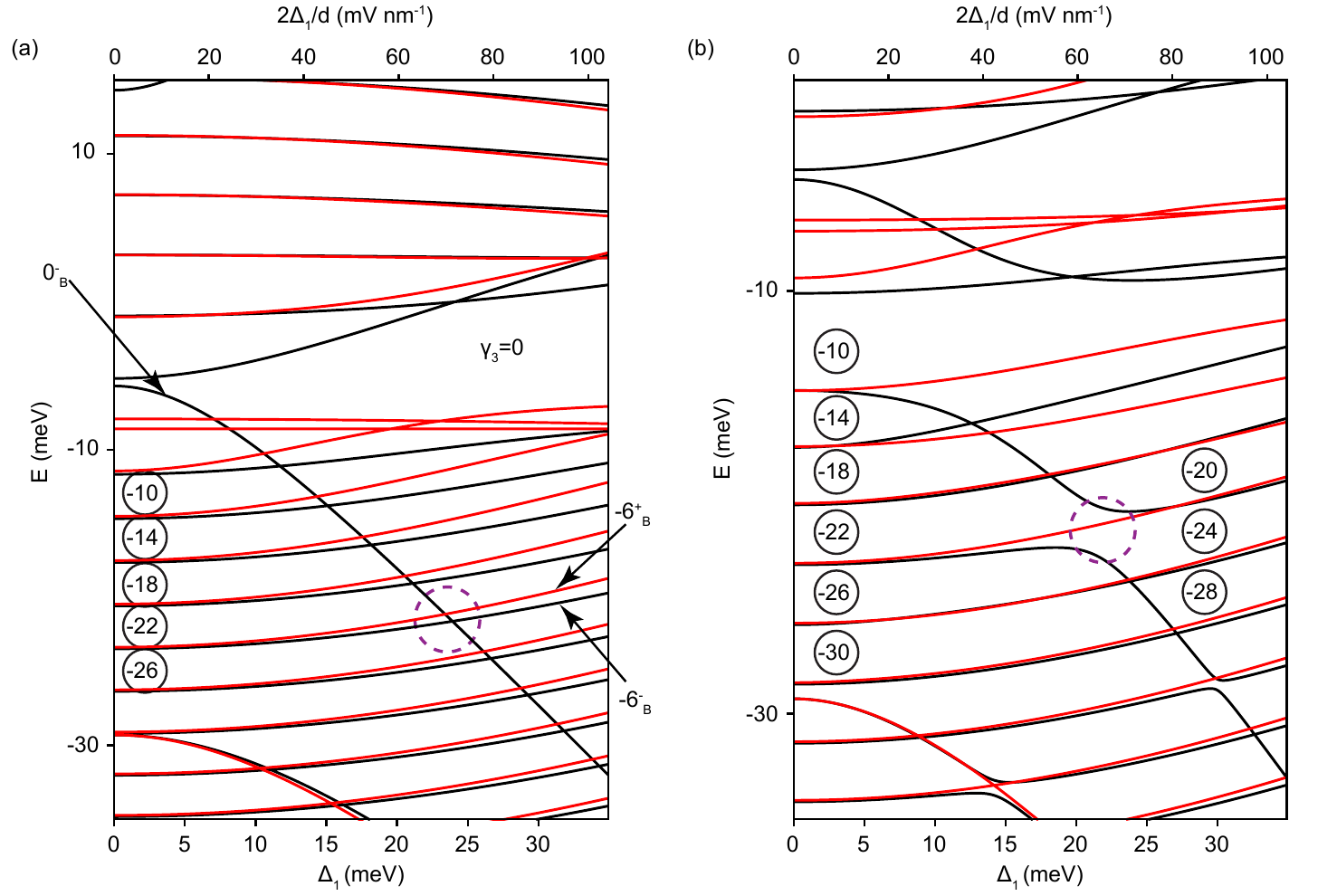}
\caption{ \label{fig:Sfig9} (a) LL energy as a function of electric field at 1.5~T for $\gamma_3$=0 showing no anticrossings. 0$_\mathrm{B}^-$ and -6$_\mathrm{B}^-$ LLs are marked which show anticrossing in presence of $\gamma_3$. \textit{K}$_-$ and 	\textit{K}$_+$ valleys are shown in black and red colours respectively. (b) Calculated LL diagram in presence of $\gamma_3$ showing the change of the marked crossing point from a crossing to an anticrossing. The circled numbers denote filling factors.}
\end{figure}

We study the evolution of LLs with electric field for some values of magnetic field. Fig.~\ref{fig:Sfig4}a and Fig.~\ref{fig:Sfig4}b show two such measurements at very low field of 0.25~T and 1.5~T respectively. Fig.~\ref{fig:Sfig5} shows a line plot of  \textit{G}$_\mathrm{xx}$ at 1.5~T along zero electric field direction. Fig.~\ref{fig:Sfig6} shows detailed measurements at 0.5~T magnetic field and the theoretically calculated DOS(\textit{E}$_\mathrm{F}$) which mimics the experimentally measured conductance. Among many features, we focussed particularly on two points in the main text. First, the clear observation that electric field  increases the separation of 0$_\mathrm{M}^+$ and 0$_\mathrm{M}^-$. This is consistent with the electric field dependence of the monolayer band gap discussed in section~\Rmnum{1} of Supplemental Material and the calculated evolution of the 0$_\mathrm{M}^+$ and 0$_\mathrm{M}^-$  LLs (Fig.~\ref{fig:Sfig7}). Second, Fig.~\ref{fig:Sfig6}c shows a faint anticrossing on hole side which can be compared to the calculated DOS(\textit{E}$_\mathrm{F}$) in Fig.~\ref{fig:Sfig6}d. Fig.~\ref{fig:Sfig8} shows the corresponding anticrossings in calculated LL energy diagram. The dashed circles mark the maximally gapped anticrossings which are observed in our experiment. We infer from theory (Fig.~\ref{fig:Sfig8}) that the anticrossing at \textit{E}$^\perp$$\sim$20~mVnm$^{-1}$ in Fig.~\ref{fig:Sfig6}c corresponds to the inter-subband anticrossings of -1$_\mathrm{M}$   with other BLG-like LLs. In contrast, the feature at \textit{E}$^\perp$$\sim$38~mVnm$^{-1}$ in Fig.~\ref{fig:Sfig6}c corresponds to the intra-subband anticrossing between  0$_\mathrm{B}^-$ with another BLG-like LL. Fig.~\ref{fig:Sfig9} shows the anticrossing at 1.5~T for which the experimental data is shown in main text. We observe an interesting line-like feature within the anticrossing marked by a dashed rectangle (Fig.3b in main text)  unlike in traditional GaAs/AlGaAs quantum well systems~\cite{Zhang_anticrossing_2006}. This is due to the existence of two nearly degenerate valleys at low \textit{B}. When both -6$_\mathrm{B}^+$ and -6$_\mathrm{B}^-$ nearly degenerate LLs cross 0$_\mathrm{B}^-$, only -6$_\mathrm{B}^-$ shows anticrossing with 0$_\mathrm{B}^-$ (Fig.~\ref{fig:Sfig9}b). As a result, -6$_\mathrm{B}^+$ goes through the anticrossing gap producing a line-like feature within the anticrossing.

\clearpage

\section{Electric field dependence of the $\nu$=0 state}

\begin{figure}[h]
\includegraphics[width=16cm]{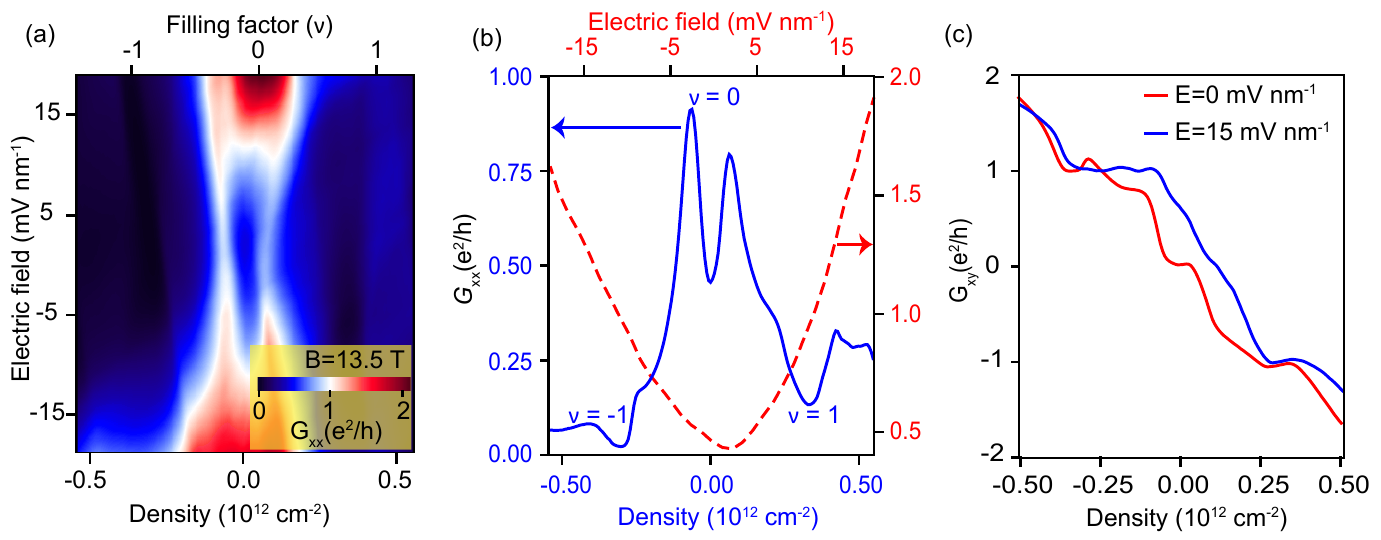}
\caption{ \label{fig:Sfig10} Dependence of $\nu$=0 state on electric field. (a) Colour plot of \textit{G}$_{\mathrm{xx}}$ showing the \textit{E}$^\perp$ dependence of the $\nu$=0 state. (b) Blue solid line is the line-cut at zero \textit{E}$^\perp$ showing a prominent dip at \textit{n}=0 which indicates the formation of the $\nu$=0 state. The red dashed line is the line-cut at \textit{n}=0 showing the disappearance of the minima with increasing  \textit{E}$^\perp$. (c) Electric field dependence of the corresponding Hall conductance (\textit{G}$_{\mathrm{xy}}$).}
\end{figure}

An independent verification of the calculated LL spectra using our band parameters can be seen from the fragility of the $\nu$=0 state in presence of a small \textit{E}$^\perp$. Fig.~\ref{fig:Sfig10}a shows that $\nu$=0 is best resolved at zero \textit{E}$^\perp$ and disappears fast with increasing \textit{E}$^\perp$. Fig.~\ref{fig:Sfig10}b shows a well-developed minima in \textit{G}$_{\mathrm{xx}}$ at zero \textit{n} and its disappearance with increasing \textit{E}$^\perp$. Fig.\ref{fig:Sfig10}c shows that $\nu$=0 Hall plateau vanishes with increasing electric field. From the Fig.~4d in main text, $\nu$=0 state at 14~T appears between 1$_\mathrm{B}^{+\downarrow}$ and 1$_\mathrm{B}^{-\uparrow}$ LLs. It also shows that \textit{E}$^\perp$ reduces the energy of 1$_\mathrm{B}^{+\downarrow}$ shrinking the $\nu$=0 gap which explains the disappearance of  $\nu$=0 with increasing \textit{E}$^\perp$. This is consistent with the electric field dependence of \textit{G}$_{\mathrm{xx}}$.

\clearpage

\section{Comparison of band gap between MLG-like bands from literature}

Here we show a comparison table (Table~\ref{Tab:tab1}) for the band gap between MLG-like bands used in literature. Theoretically, the band gap between MLG-like bands can be calculated from the band parameters as  \textit{E}$_\mathrm{g}$=$\delta$+$\frac{\gamma_2}{2}$-$\frac{\gamma_5}{2}$. Our experiment suggests that the band gap is much smaller than it was assumed in most previous studies. We find the band gap to be \textit{E}$_\mathrm{g}$$\sim$1~meV (see Fig.~\ref{fig:Sfig12} for the comparison between theoretical and experimental data) whereas previous studies assumed a large distribution of values. Our work provides a direct way to estimate this band gap by simply keeping track of the LLs originated from the band edges.

\begin{table*}[h]
\caption{ \label{Tab:tab1} \textbf{Comparison of band gap between MLG-like bands from literature}}

 \begin{tabular}{|c|c|c|c|c|}
  \hline
 Reference & $\delta$ (meV) & $\gamma_2$ (meV) & $\gamma_5$ (meV) & \textit{E}$_\mathrm{g}$=($\delta$+$\frac{\gamma_2}{2}$-$\frac{\gamma_5}{2}$) (meV) \\ \hline

This work &  20 &  -20 &  18 &  1 \\ \hline
Campos, Leonardo C., et al. PRL (2016) [\cite{campos_landau_2016_s}]  &  15 &  -18 &  10 &  1 \\ \hline
Stepanov, Petr, et al. PRL (2016) [\cite{stepanov_tunable_2016_s}] &  27 &  -32 &  10 &  6 \\ \hline
Shimazaki, Yuya, et al. [\cite{shimazaki_landau_2016_s}] &  46 &  -28 &  50 &  7 \\
                             &  34 &  -28 &  50 &  -5 \\ \hline
Taychatanapat, Thiti, et al.  Nat. Physics (2011) [\cite{taychatanapat_quantum_2011_s}] &  46 &  -28 &  50 &  7 \\ \hline
Asakawa, Yuta, et al. PRL (2017) [\cite{inplane_B_s}] &  14.3 &  -23.7 &  6 &  -0.55 \\ \hline

\end{tabular}
\end{table*}

\clearpage

\section{Non-uniform electric field and comparison of $\Delta_2$ value}

Here we show a comparison table (Table~\ref{Tab:tab2}) for the value of $\Delta_2$ used in literature. We emphasize that $\Delta_2$ plays a very important role and can not be neglected in the following two cases. First, for small filling factors where the order of LLs depends very sensitively on the exact value of $\Delta_2$. This is shown to be crucial to explain the LL crossing pattern in the main text. Second, the assumption of uniform electric field breaks down dramatically in a regime when $\Delta_1$ value is comparable to $\Delta_2$. For example, the data presented in the main text (Fig.4a and Fig.4c) shows that all low filling factor crossings occur for  $\Delta_1 \leq$10~meV. In this regime, even for the maximum $\Delta_1$=10~meV we show that the perpendicular electric field is very non-uniform. From the potential energy distribution in the three layers of ABA-stacked TLG shown in main text (Fig.1c), the electric field between the top and middle layers is \textit{E}$^\perp_{\mathrm{top}}$=$\frac{(\Delta_1+\Delta_2)-(-2\Delta_2)}{d/2}$=68.3~mVnm$^{-1}$ where $\Delta_2$=4.3~meV and d=0.67~nm. In contrast the electric field between the middle and bottom layers is \textit{E}$^\perp_{\mathrm{bottom}}$=$\frac{(-2\Delta_2)-(-\Delta_1+\Delta_2)}{d/2}$=-8.6~mVnm$^{-1}$. We note that not only the values of electric fields are extremely different, their directions are also opposite. This regime can not be described in terms of an average electric field \textit{E}$^\perp_{\mathrm{av}}$=$\frac{2\Delta_1}{d}$=29.8~mVnm$^{-1}$. This shows the importance of the $\Delta_2$ and our work shows a systematic way to determine this important parameter.

\begin{table*}[h]
\caption{ \label{Tab:tab2} \textbf{Comparison of the $\Delta_2$ value used in literature}}

 \begin{tabular}{|c|c|c|c|c|c|c|c|c|c|c|c|}
  \hline
 Reference & $\Delta_2$ (meV) \\ \hline

 This work &  4.3 \\ \hline

Campos, Leonardo C., et al. PRL (2016) [\cite{campos_landau_2016_s}] &   1.8 and 5.7 \\ \hline

Stepanov, Petr, et al. PRL (2016) [\cite{stepanov_tunable_2016_s}] &  1.8 \\ \hline

Shimazaki, Yuya, et al. [\cite{shimazaki_landau_2016_s}] &  0 \\ \hline
Taychatanapat, Thiti, et al.  Nat. Physics (2011) [\cite{taychatanapat_quantum_2011_s}] &  0 \\ \hline
Asakawa, Yuta, et al. PRL (2017) [\cite{inplane_B_s}] &  0 \\ \hline
\end{tabular}

\end{table*}

\clearpage

\section{Fractional quantum Hall phases in ABA-stacked trilayer graphene}

\begin{figure}[h]
\includegraphics[width=12cm]{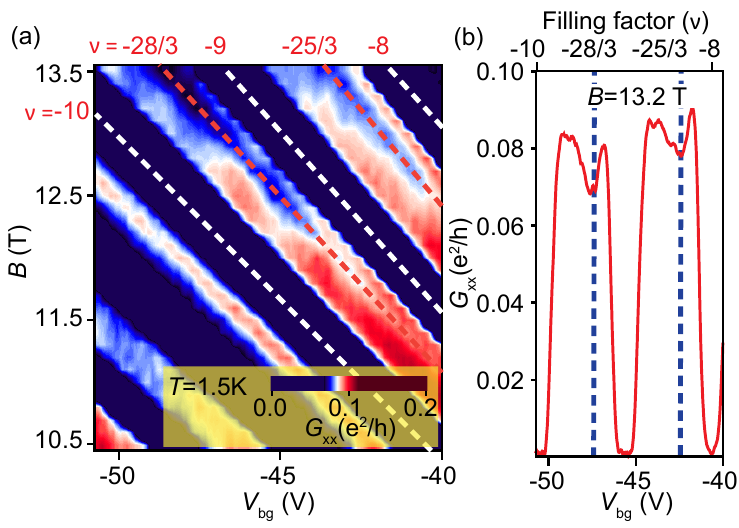}
\caption{ \label{fig:Sfig11} (a) LL fan diagram showing the development of two FQH states at high \textit{B}. (b) Line-cut at 13.2~T showing two clear dips in \textit{G}$_{\mathrm{xx}}$ at $\nu$=-25/3 and -28/3.}
\end{figure}

Though all the LL crossing physics we have discussed so far can be understood using single particle picture, at high \textit{B} electronic interaction becomes important~\cite{jang_stacking_2015_s,datta_strong_2017_s} which is evident in some partially  developed fractional quantum Hall (FQH) states~\cite{Apalkov_2012_s,dean_multicomponent_2011_s}. We report two partially developed FQH phases in our experiment at much lower \textit{B} compared to the previous study~\cite{stepanov_tunable_2016_s} where \textit{B} up to 45~T was used. Fig.~\ref{fig:Sfig11}a shows the appearance of two additional minima in \textit{G}$_{\mathrm{xx}}$ at $\nu$=-25/3 and -28/3. The two minima become discernible with increasing \textit{B} from 11~T.  According to the calculated LL diagram (Fig.2 in main text), $\nu$=-6 corresponds to completely filled 0$_\mathrm{B}$  and 1$_\mathrm{B}$ LLs on the hole side. So, from filling factor consideration, it is clear that these FQH states develop from -2$_\mathrm{B}$ BLG-like LL. Observation of these \textit{E}$^\perp$ tunable FQH states~\cite{maher2014tunable_s} at low \textit{B} opens up a new avenue to study recently predicted~\cite{Jain_graphene_s} non-Abelian parton states in multilayer graphene.

\clearpage

\section{Details of the theoretical calculation}

We calculate the Landau level spectra of ABA-stacked trilayer graphene using a tight binding
model. The LLs are broadened by a phenomenological parameter to obtain a Lorentzian DOS
peaked around the LL energy. We then calculate electron/hole density integrating
this DOS up to the chemical potential ($E_\mathrm{F}$). We  plot the DOS at the Fermi level
as a function of density and electric field. The DOS at the Fermi level shows all the features
observed in the behaviour of longitudinal conductance ($G_{\mathrm{xx}}$) in the experiment. We vary the
tight binding parameters and for each set of parameters, we read off the $E^\perp$ where the
levels cross. These values are matched with experimental data to constrain the tight binding
parameters.

In presence of electric field, the reflection symmetry between the top and bottom layer
is broken in a trilayer graphene and therefore the monolayer-like and bilayer-like bands
of trilayer graphene get hybridized.
In the rotated basis $\left\{ \frac{A_1-A_3}{\sqrt{2}}, \frac{B_1-B_3}{\sqrt{2}},
\frac{A_1+A_3}{\sqrt{2}}, B_2, A_2, \frac{B_1+B_3}{\sqrt{2}} \right\}$ the
trilayer graphene Hamiltonian can be written as \cite{serbyn_new_2013_s}

\begingroup
\footnotesize

\begin{equation}
H_\text{TLG} = \left(\begin{array}{cccccc}
\Delta_2-\frac{\gamma_2}{2} & v_0\pi^\dagger & \Delta_1 & 0 & 0 & 0 \\
v_0\pi & \Delta_2+\delta-\frac{\gamma_5}{2} & 0 & 0 & 0 & \Delta_1 \\
\Delta_1 & 0 & \Delta_2+\frac{\gamma_2}{2} & \sqrt{2}v_3\pi
& -\sqrt{2}v_4\pi^\dagger & v_0\pi^\dagger \\
0 & 0 & \sqrt{2}v_3\pi^\dagger & -2\Delta_2 & v_0\pi & -\sqrt{2}v_4\pi \\
0 & 0 & -\sqrt{2}v_4\pi & v_0\pi^\dagger & \delta-2\Delta_2 & \sqrt{2}\gamma_1 \\
0 & \Delta_1 & v_0\pi & -\sqrt{2}v_4\pi^\dagger & \sqrt{2}\gamma_1
& \Delta_2+\delta+\frac{\gamma_5}{2}
\end{array}\right)
\end{equation}

\endgroup

where $\Delta_1=-e.\frac{U_1-U_3}{2}$ and $\Delta_2 = -e.\frac{U_1-2U_2+U_3}{6}$, with the
potential of the layer $i$ is given by $U_i$. $\Delta_1$ is related to the average $E^\perp$
by \textit{E}$^\perp_{\mathrm{av}}$=$\frac{2 \Delta_1}{(e)d}$, while $\Delta_2$ controls the asymmetry of the electric
field between the layers. The band velocity $v_i (i=0,3,4)$ is related to the tight
binding parameter ($\gamma_i$) by $v_i \hbar = \frac{\sqrt{3}}{2}a\gamma_i$, where
$a$ is the lattice constant. $\pi$ is defined in different valleys as $\pi=\zeta k_x+ik_y$
with $\zeta=\pm 1$ for $K_+$ and $K_-$ valley respectively.
The action of $\pi$ and $\pi^\dagger$ on harmonic oscillator states
in different valleys are given by \cite{serbyn_new_2013_s}
\begin{eqnarray}
 K_+ : \pi|n\rangle &=& \frac{i\hbar}{l_B}\sqrt{2(n+1)}|n+1\rangle  \nonumber\\
 \pi^\dagger|n\rangle &=& -\frac{i\hbar}{l_B}\sqrt{2n}|n-1\rangle \nonumber\\
%%%%%%%%
 K_- : \pi|n\rangle &=& \frac{i\hbar}{l_B}\sqrt{2n}|n-1\rangle  \nonumber\\
 \pi^\dagger|n\rangle &=& -\frac{i\hbar}{l_B}\sqrt{2(n+1)}|n+1\rangle \nonumber\\
\label{operators}
\end{eqnarray}
where, magnetic length $l_B = \sqrt{\hbar/eB}$. They satisfy the commutation relation
$[\pi,\pi^\dagger]=1$ in each valley.

The Landau levels for monolayer-like and bilayer-like blocks are hybridized by the
off-diagonal matrix elements ($\Delta_1$). In absence of $\Delta_1$, the trilayer
graphene Hamiltonian can be block diagonalized into monolayer-like and bilayer-like
sectors i.e. $H_\text{TLG}=H_\text{MLG} \oplus H_\text{BLG}$.
Let us look at $K_+$ valley first. $H_\text{MLG}$ can be diagonalized in an eigenbasis
$\big\{ |n-1\rangle , |n\rangle \big\}$,
where $n$ runs from 0 to $\infty$. Similarly in absence of $\gamma_3$,
$H_\text{BLG}$ can be diagonalized in an eigenbasis
$\big\{ |m-2\rangle , |m\rangle , |m-1\rangle , |m-1\rangle \big\}$
, where $m$ runs from 0 to $\infty$. It is evident that in presence of $\Delta_1$,
$(n+1)$-th monolayer-like Landau level gets coupled with $m$-th bilayer-like Landau level
(where both $n$ and $m$ runs from 0 to $\infty$).
In $K_+$ valley $H_\text{TLG}$ can be block diagonalized in $6\times6$ sectors (one for each
$n$) if we choose $m=n+1$ as following.

\begingroup
\footnotesize

\begin{equation}
\left(\begin{array}{cccccc}
\Delta_2-\frac{\gamma_2}{2} & v_0\pi^\dagger & \Delta_1 & 0 & 0 & 0 \\
v_0\pi & \Delta_2 + \delta - \frac{\gamma_5}{2} & 0 & 0 & 0 & \Delta_1 \\
\Delta_1 & 0 & \Delta_2 + \frac{\gamma_2}{2} & 0
& -\sqrt{2}v_4\pi^\dagger & v_0\pi^\dagger \\
0 & 0 & 0 & -2\Delta_2 & v_0\pi & -\sqrt{2}v_4\pi \\
0 & 0 & -\sqrt{2}v_4\pi & v_0\pi^\dagger & \delta-2\Delta_2 & \sqrt{2}\gamma_1 \\
0 & \Delta_1 & v_0\pi & -\sqrt{2}v_4\pi^\dagger & \sqrt{2}\gamma_1
& \Delta_2+\delta+\frac{\gamma_5}{2}
\end{array}\right)
\left(\begin{array}{c}
c_1|n-1\rangle \\ c_2|n\rangle \\ c_3|n-1\rangle \\ c_4|n+1\rangle \\ c_5|n\rangle
\\ c_6|n\rangle
\end{array}\right)
=
\left(\begin{array}{c}
c_1'|n-1\rangle \\ c_2'|n\rangle \\ c_3'|n-1\rangle \\ c_4'|n+1\rangle \\ c_5'|n\rangle
\\ c_6'|n\rangle
\end{array}\right)
\end{equation}

\endgroup

Similarly, in $K_-$ valley $H_\text{MLG}$ can be diagonalized in an eigenbasis
$\big\{ |n\rangle , |n-1\rangle \big\}$ and $H_\text{BLG}$ can be diagonalized in an
eigenbasis $\big\{ |m\rangle , |m-2\rangle , |m-1\rangle , |m-1\rangle \big\}$
(in absence of $\gamma_3$), where $n$ and $m$ both run from 0 to $\infty$.
In presence of $\Delta_1$, $n$-th monolayer-like Landau level gets coupled with
$m$-th bilayer-like Landau level (where both $n$ and $m$ runs from 0 to $\infty$).
In $K_-$ valley $H_\text{TLG}$ can be block diagonalized in $6\times6$ sectors if we
choose $m=n$ as following.

\begingroup
\footnotesize

\begin{equation}
\left(\begin{array}{cccccc}
\Delta_2-\frac{\gamma_2}{2} & v_0\pi^\dagger & \Delta_1 & 0 & 0 & 0 \\
v_0\pi & \Delta_2 + \delta - \frac{\gamma_5}{2} & 0 & 0 & 0 & \Delta_1 \\
\Delta_1 & 0 & \Delta_2 + \frac{\gamma_2}{2} & 0
& -\sqrt{2}v_4\pi^\dagger & v_0\pi^\dagger \\
0 & 0 & 0 & -2\Delta_2 & v_0\pi & -\sqrt{2}v_4\pi \\
0 & 0 & -\sqrt{2}v_4\pi & v_0\pi^\dagger & \delta-2\Delta_2 & \sqrt{2}\gamma_1 \\
0 & \Delta_1 & v_0\pi & -\sqrt{2}v_4\pi^\dagger & \sqrt{2}\gamma_1
& \Delta_2+\delta+\frac{\gamma_5}{2}
\end{array}\right)
\left(\begin{array}{c}
c_1|n\rangle \\ c_2|n-1\rangle \\ c_3|n\rangle \\ c_4|n-2\rangle \\ c_5|n-1\rangle
\\ c_6|n-1\rangle
\end{array}\right)
=
\left(\begin{array}{c}
c_1'|n\rangle \\ c_2'|n-1\rangle \\ c_3'|n\rangle \\ c_4'|n-2\rangle \\ c_5'|n-1\rangle
\\ c_6'|n-1\rangle
\end{array}\right)
\end{equation}

\endgroup

In presence of $\gamma_3$, the rotational symmetry gets broken down to $C_3$ and hence
states with quantum numbers differing by 3 couple to each other. This, in principle makes
the diagonalization problem infinite dimensional. However we put a cut off on the matrix
size and diagonalize large but finite matrices to obtain the spectra. We find that
$N_\text{max}\sim 100$ is sufficient to obtain the dispersion of low lying Landau levels
to our desired accuracy.

The Landau levels of trilayer graphene Hamiltonian get broadened in presence of scattering
and the density of states of each Landau level (with energy $E$) can be approximated by a
Lorentzian
\begin{equation}
 \text{DOS}(E) = \frac{1}{2\pi l_B^2}\frac{1}{\pi}
 \left( \frac{\Gamma}{E^2+\Gamma^2} \right)
\end{equation}
where the effect of disorder has been incorporated in $\Gamma$, which is related to the
relaxation time ($\tau$) of scattering as $\Gamma\sim {(1/\tau)}^{1/2}$ \cite{AndoOSC_s}.

We calculate the density of states at Fermi energy (DOS$(E_\mathrm{F})$) as a function of
total density $\textit{n}$ (integrating DOS($E$) up to Fermi energy $E_\mathrm{F}$) and $\Delta_1$ to match
the experimental data. We have found good theoretical match with the experimental data with
a choice of $\Gamma$=1.2~meV on the electron-side and $\Gamma$=0.85~meV on
the hole-side of the dispersion. The choice of different $\Gamma$ is supported by the fact
that mobility on the hole-side is more than on the  electron-side.

\clearpage

\section{Details of the fitting to determine band parameters}

The detailed band structure of ABA-stacked TLG has 9 parameters (hopping amplitude and potentials) which affect the Landau level spectrum. The details of the meaning of these parameters are given in the main text. There are 3 hopping amplitudes, $\gamma_0$=3.1~eV, $\gamma_1$=0.39~eV, $\gamma_3$=315~meV, which are  larger than other parameters and are widely and consistently reported in the literature. While fitting band parameters we take these values as given. The  potential difference between the first and the third layer, $\Delta_1$ is related to the electric field, which is measured in the experiments. This leaves the following parameters to be determined: $\gamma_2$, $\gamma_4$, $\gamma_5$, $\delta$ and $\Delta_2$.  Constraining large band parameters using experimental low LL index crossings are not effective which is evident in the large range of $\gamma_4$ (40-140~meV) in the previous study~\cite{campos_landau_2016_s}. Here we use previously reported value of $\gamma_4$ =120 meV and do not try to constrain it using our data.

In our low magnetic field (1.5 T) data the LLs are not fully developed and their crossings form an almost continuous curve in the electric field-density plane. While this gives a large number of crossings to fit, the LL spectrum is insensitive to the large parameters in this limit and the crossing pattern is sensitive to $\gamma_2$, $\gamma_5$, $\delta$ while being less sensitive to $\Delta_2$. We will later use high field data at 8 and 14 T  to constrain $\Delta_2$ tightly.

\subsection{Determination of $\gamma_2$, $\gamma_5$ and $\delta$}

We use all the 12 LL crossing points (due to 0$_\mathrm{B}^-$ and -1$_\mathrm{M}$ LLs) visible in Fig.3b main text (at 1.5 T) to determine the band parameters $\gamma_2$, $\gamma_5$ and $\delta$. We start with the bulk graphite values $\gamma_2$=-20~meV, $\gamma_5$=38~meV, $\delta$=8~meV and vary each of the band parameters around them till we minimize the fitting error in electric field at the correct filling factors. We define the fitting error = $\frac{1}{12}$$\sqrt{\sum_{i=0}^{12}(\mathrm{E}_i^{Exp}-\mathrm{E}_i^{Th})^2}$ where $\mathrm{E}_i^{Exp}$ and $\mathrm{E}_i^{Th}$ are the experimental and theoretical electric fields at the i\textquotesingle th LL crossing point respectively. The minimum fitting error in $\Delta_1$ at 1.5~T using the quoted band parameters in the main text is $\sim$0.8~meV.  The filling factor range in this fitting is -62 to -14. Fitting this data  keeping $\Delta_2$ as a parameter gives us a range 3-5.5~meV. Since $\Delta_2$ is small compared to the energy of the LLs in this high filling factor regime, it is not possible to further constrain it with better accuracy. So, we determine the final value of $\Delta_2$ from the crossings within zeroth LLs at high magnetic fields described below.

\begin{figure}[h]
\includegraphics[width=15.5cm]{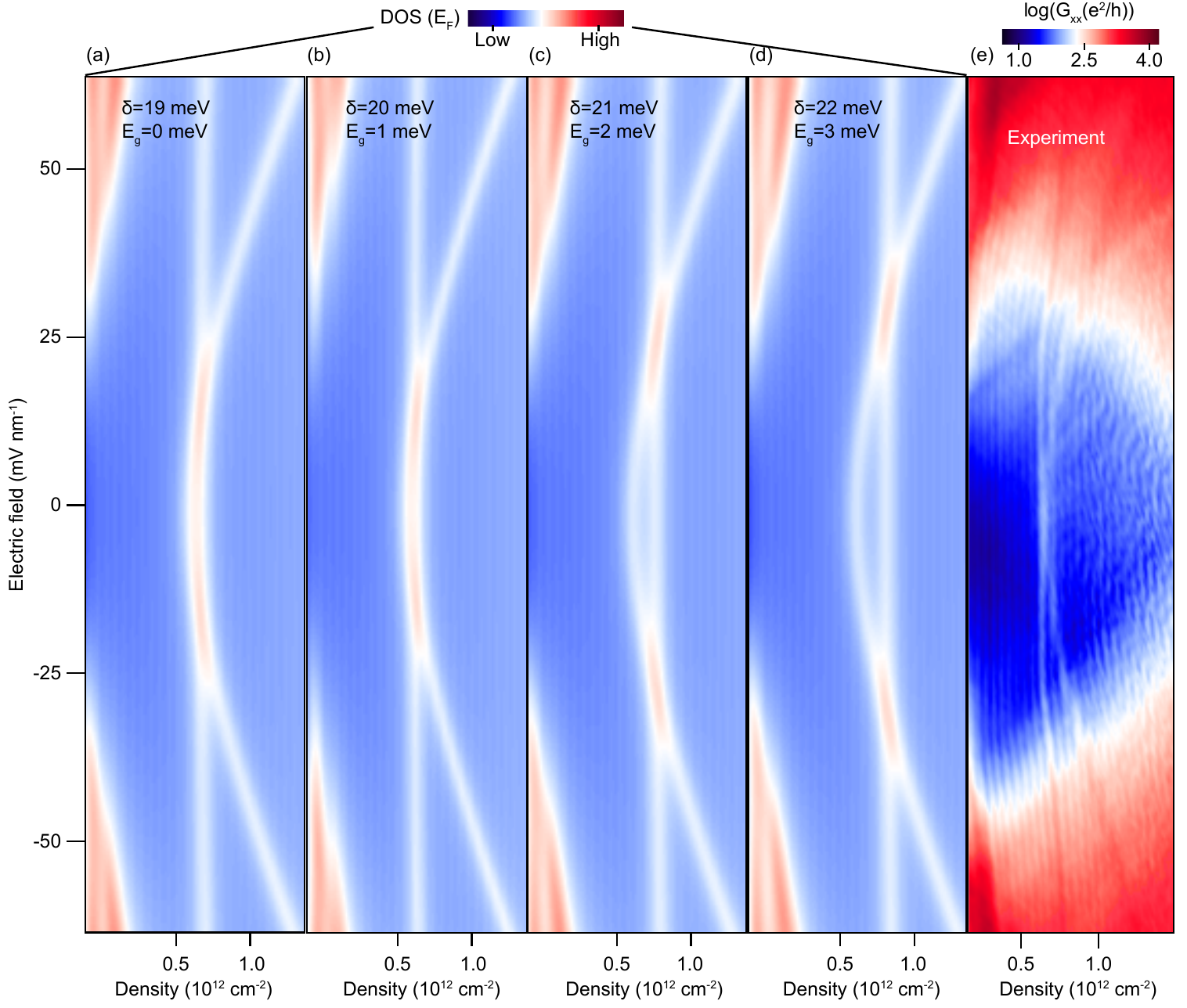}
\caption{ \label{fig:Sfig12} (a)-(d) Numerically calculated DOS at 0.5~T varying $\delta$=19-22~meV which changes the valley gap of 0$_\mathrm{M}$ LL 0-3~meV. (e) Experimental G$_\mathrm{xx}$ at 0.5~T.}
\end{figure}

Additionally, we find the valley gap of 0$_\mathrm{M}$ Landau level is
\textit{E}$_\mathrm{g}$=$\delta$+$\frac{\gamma_2}{2}$-$\frac{\gamma_5}{2}$ $\sim$1~meV as discussed in Supplemental Material~\Rmnum{6} . This comes from comparing the experimental data with the calculated DOS (Fig.~\ref{fig:Sfig12}). From our experiment, we see that the valley splitting of the 0$_\mathrm{M}$ LL cannot be resolved at zero electric field meaning the valley splitting to be $\sim$1~meV for a disorder of $\sim$1~meV (calculated from Dingle plot). Calculated DOS shows that if the valley gap is more than 1~meV, it reflects immediately in the density axis and becomes resolvable. We show the calculated DOS (Fig.~\ref{fig:Sfig12}) for a few other valley gaps clearly showing that the valley gap has to be $\sim$1~meV. This additional experimental constraint allows us to estimate $\gamma_2$, $\gamma_5$ and $\delta$ more accurately.

\subsection{Determination of $\Delta_2$}

The exact value of $\Delta_2$ becomes important in the zeroth LLs because the magnitude of $\Delta_2$ is comparable to the separation between this manifold. This is clear from the approximated analytic energy of these LLs at zero electric field limit:

  \begin{equation}
     E(0_\mathrm{B}^+)=-2\Delta_2
     \label{equn:equn1}
    \end{equation}

     \begin{equation}
     E(0_\mathrm{B}^-)=-\frac{|\gamma_2|}{2}+\Delta_2
     \label{equn:equn2}
    \end{equation}

    \begin{equation}
     E(1_\mathrm{B}^+)=-2\Delta_2 + \xi (\frac{\gamma_5}{2}+\delta+\Delta_2)
     \label{equn:equn3}
    \end{equation}

   \begin{equation}
     E(1_\mathrm{B}^-)=-\frac{|\gamma_2|}{2}+\Delta_2 + \xi (\delta-2\Delta_2)
     \label{equn:equn4}
    \end{equation}

\begin{figure}[h]
\includegraphics[width=15.5cm]{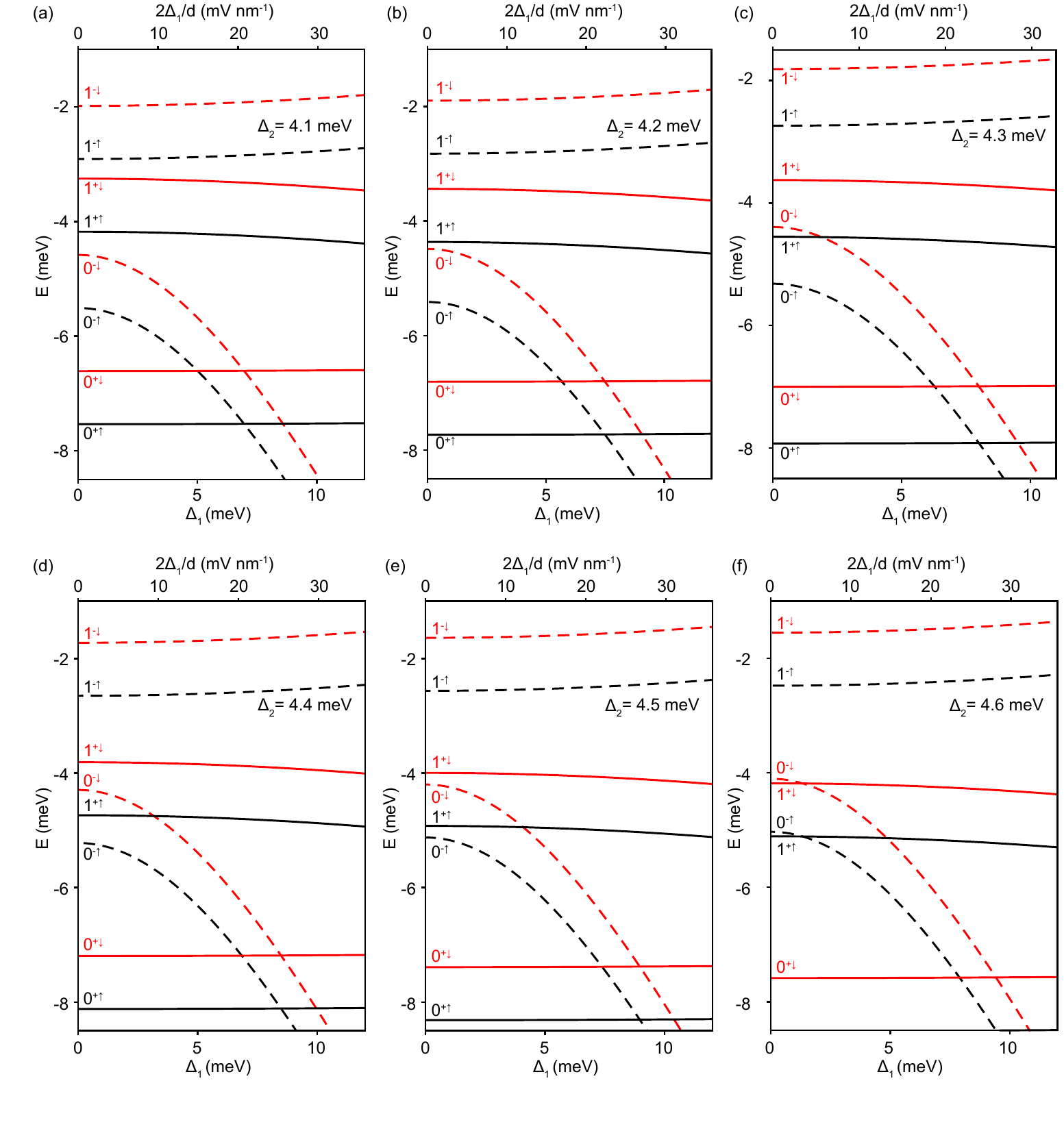}
\caption{ \label{fig:Sfig13} (a)-(f) Electric field dependence of zeroth LLs at 8~T for $\Delta_2$=4.1-4.6~meV.}
\end{figure}

\begin{figure}[h]
\includegraphics[width=15.5cm]{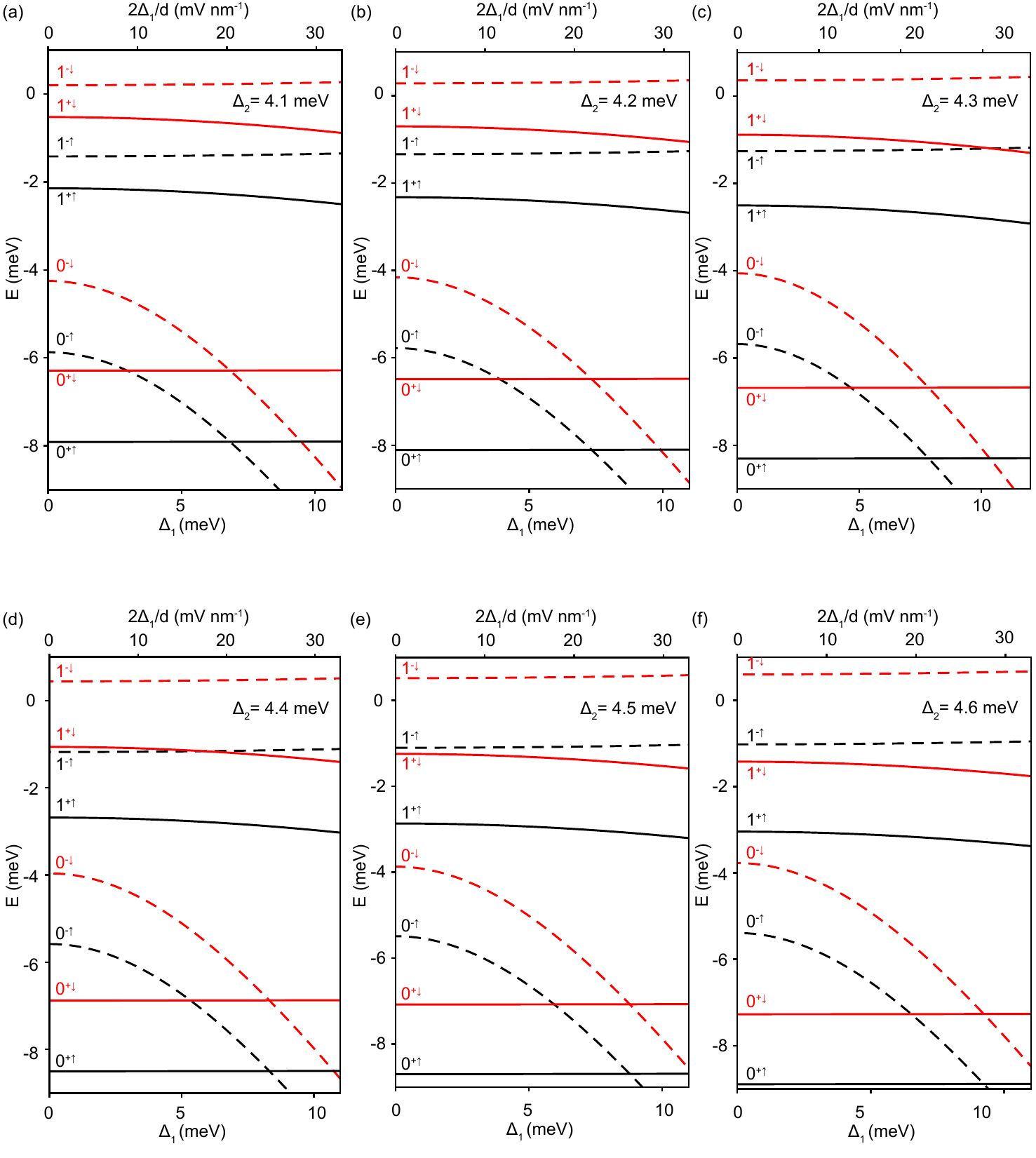}
\caption{ \label{fig:Sfig14} (a)-(f) Electric field dependence of zeroth LLs at 14~T for $\Delta_2$=4.1-4.6~meV.}
\end{figure}

Here $\xi$ is a dimensionless parameter given by $\xi$=$\frac{\hbar \omega_c}{\sqrt{2}\gamma_1}$=$\frac{\hbar e B v^2}{\gamma_1^2}$=4.35$\times 10^{-3}$\textit{B}. We clearly see that the energy of these LLs depend directly on $\delta$, $\gamma_2$, $\gamma_5$ and $\Delta_2$. So, the LL crossings within zeroth LLs provide a precise cross check on the band parameters determined above. We note that the valley gap of both 0$_\mathrm{B}$ and 1$_\mathrm{B}$ LLs (at small electric and magnetic field) is $\sim\frac{|\gamma_2|}{2}-3\Delta_2$ which is very sensitive on $\Delta_2$ as it comes with a factor of 3. So, the valley gap can be negative or positive depending on the value of $\Delta_2$; this determines the sequence of the LLs and hence has a bearing on the LL crossing pattern. Fig.~\ref{fig:Sfig13} and Fig.~\ref{fig:Sfig14} show the calculated LL diagrams for $\Delta_2$ values 4.1-4.6~meV at 8~T and 14~T respectively. This clearly shows the crossing pattern is very sensitive to $\Delta_2$. Experimental data provides the following constraints on the crossing pattern:

\begin{enumerate}
\item At 8~T, there is one LL crossing at each $\nu$=-2,-3 and -5, but at $\nu$=-4 there are two LL crossings (Fig.4a).
\item At 14~T, $\nu$=-2 crossing disappears (Fig.4c).
\item LL crossings at $\nu$=-3 and $\nu$=-5 are almost at the same electric field (both at 8~T and 14~T, Fig.4a and Fig.4c).
\item $\nu$=0 gap at 14~T closes with the electric field (Fig.\ref{fig:Sfig10}).
\end{enumerate}

Only $\Delta_2$=4.3-4.4~meV can explain the crossing pattern observed at 8~T and 14~T simultaneously. Other values of $\Delta_2$ result in incorrect crossing pattern.

\begin{thebibliography}{49}
\bibitem [{foo({\natexlab{b}})}]{footnote2}%
  \BibitemOpen
  \href@noop {} {\bibfield  {journal} {\bibinfo  {journal} {Note that
  $\gamma_2$ is already tightly constrained by matching our low magnetic field
  data with theoretical calculations}\ } ({\natexlab{b}})}\BibitemShut
  {NoStop}%
\end{thebibliography}
\end{document}